\documentclass{aa}  

\usepackage{graphicx}
\usepackage{booktabs}

\usepackage{txfonts}
\usepackage{longtable}
\usepackage{multicol}
\newcommand{\ha}{H$\alpha$}

\newcommand{\htwo}{H$_2$}

\newcommand{\feii}{[\ion{Fe}{ii}]}
\newcommand{\sii}{[\ion{S}{ii}]}

\newcommand{\caii}{[\ion{Ca}{ii}]}
\newcommand{\oii}{[\ion{O}{ii}]}
\newcommand{\oi}{[\ion{O}{i}]}

\newcommand{\nii}{[\ion{N}{ii}]}
\newcommand{\niii}{[\ion{Ni}{ii}]}
\newcommand{\feiii}{[\ion{Fe}{iii}]}
\newcommand{\crii}{[\ion{Cr}{ii}]}

\newcommand{\kms}{km\,s$^{-1}$}
\newcommand{\um}{$\mu$m}

\newcommand{\cmt}{cm$^{-3}$}

\newcommand{\cmu}{cm$^{-1}$}

\begin{document}

   \title{The small-scale HH34~IRS jet as seen by X-shooter
   \thanks{Based on Observations collected with X-shooter at the Very Large Telescope on Cerro Paranal (Chile), operated by the European Southern Observatory (ESO). Program ID: 090.C-0606.}}

   \author{B. Nisini
          \inst{1}
          \and
           T. Giannini\inst{1}
          \and S. Antoniucci\inst{1} 
          \and J.M. Alcal\'a\inst{2}
          \and F. Bacciotti\inst{3}
          \and L. Podio\inst{3}
          }

   \institute{INAF - Osservatorio Astronomico di Roma, Via di Frascati 33, 00078 Monte Porzio Catone, Italy\\
   \email{brunella.nisini@oa-roma.inaf.it} 
   \and INAF-Osservatorio Astronomico di Capodimonte, via Moiariello, 16, 80131, Napoli, Italy 
   \and INAF-Osservatorio Astrofisico di Arcetri, Largo E. Fermi 5, 50125 Firenze, Italy
             }

   \date{Received ; accepted }


  \abstract
   {Very little information has 
   been so far gathered on atomic jets from young
   embedded low mass sources (class I stars), especially in the inner jet 
   region.}
{We exploit multiwave spectroscopic observations
   to infer physical conditions of the inner region of HH34~IRS, a prototypical class I jet.}
   {We use a deep X-shooter spectrum ($\lambda \sim$ 350-2300 nm, $R$ between $\sim$8000 and $\sim$18000)
  detecting lines from E$_{up} \sim$ 8000 to $\sim$ 31\,000 \cmu . 
 Statistical 
   equilibrium and ionization models are adopted to derive the jet main physical parameters. 
    }
   {We derive the physical conditions for the extended high velocity jet (HVC, V$_{r} \sim $ $-$100\kms\,) and
 for the low velocity and compact gas  (LVC, V$_{r} \sim $ $-$20-50\kms). 
 At the jet base ($<$ 200 AU) the HVC is mostly neutral (x$_e <$ 0.1)
 and very dense ($n_H >$ 5\,10$^5$ \cmt).
The LVC gas has the same density and A$_V$ as the HVC, but it is a factor of two colder. Iron abundance in the HVC is close to solar, while it is 3 times subsolar in the LVC, suggesting an origin of the LVC gas from a dusty disk. 
We infer a \.M$_{jet}$ of $>$5\,10$^{-7}$ M$_{\odot}$\,yr$^{-1}$. We find that the relationships between accretion luminosity and line luminosity derived for T Tauri stars cannot be directly extended to class I sources surrounded by reflection nebulae since the permitted lines are seen through scattered light. An accretion rate of $\sim$ 7\,10$^{-6}$ M$_{\odot}$\,yr$^{-1}$
is instead derived from an empirical relationship connecting \.M$_{acc}$ 
with $L$(\oi)(HVC).
   }
   {The HH34~IRS jet shares many 
   properties with jets from active CTTs stars. It is however denser 
   as a consequence of the larger \.M$_{jet}$. These findings suggest that the acceleration and 
   excitation mechanisms in jets are not influenced by evolution and are
   similar in CTTs and in still embedded and highly accreting sources.
}   


   \maketitle
%

\section{Introduction}

High velocity outflows from young stellar objects (YSOs) are intimately
related to the accretion process of star formation. It has been suggested
that outflows from accreting YSOs 
have a fundamental role in removing
angular momentum from circumstellar disk  material allowing infalling matter to 
accrete onto the central star. 
Hence, determining the nature of the mechanism responsible for the formation 
of jets in young stars is critical in order to understand the underlying accretion. 
The details of this coupled accretion-ejection mechanism are still  poorly
understood, although magnetic fields  almost certainly play a
central role in this process. 
Whether originating from the stellar object, from the inner disk edge,
from a large section of the accreting disk, or from a combination of the 
above  (e.g., Ferreira et al. 2006, Shang, Romanova et al. 2009), 
the jet launching region 
is predicted to be located at fractions of AU from the central source, thus preventing
a direct observational view with current optical/near-infrared telescope technology.
However, constraints on the jet launching mechanism can be inferred from
observations of the outflows within a few arcsec of the central star where the jet is
expected to be still largely unaffected by ambient gas. 
In particular, significant effort has been  devoted to inferring 
kinematic and excitation conditions in these small-scale micro-jets 
from classical T Tauri (CTT) stars, where the original circumstellar 
envelope has  already been cleared out and the star--disk
interaction region is exposed to a direct observational view
(see, e.g., Ray et al. 2007, Frank et al. 2013 and references therein).

Investigation of jets from 
younger and more embedded objects (the so-called class I objects) is 
more challenging, however,  owing to the high extinction close to the source that limits
a direct view of the jet base. 
In class I sources, the jet base has been studied mainly in the IR as these micro-jets
are particularly bright in \feii\, forbidden and H$_2$ ro-vibrational 
emission lines (Davis et al. 2003, Davis et al. 2011, Garcia Lopez et al., 2008, 2010). Observations of 
\feii\, lines indicate that within the first 300 AU, class I jets
have a kinematical behavior similar to the atomic jets of 
CTT stars. They share, in particular, a similar forbidden emission line region (FEL)
consisting of a high velocity extended component (HVC) 
and a compact component close to systemic velocity (LVC)
(e.g., Hartigan et al. 1995, Davis et al. 2003, Takami et al. 2006). 
However, although \feii\, IR line ratios are used to
measure the jet electron density, no stringent constraints on the full set
of physical parameters, including temperature and ionization fraction, 
can be derived with only the IR lines. This   has limited 
a more quantitative assessment about the influence of evolutionary effects 
on the physical and dynamical properties of the class I micro-jets. 

In order to address the above issue, we report here deep observations of the class I 
source HH34~IRS obtained with the 
X-shooter instrument on VLT (Vernet et al. 2011). The X-shooter simultaneously covers
the spectral range between 0.3 to 2.4$\mu$m with a spectral resolution high enough
(up to $\sim$ 17\kms) to resolve the main jet kinematical components. 
Such observations can  therefore be very effective in probing the excitation structure of the 
jet at small scale using a wide range of diagnostic lines (Bacciotti et al. 2011,  Ellerbroek et al. 2013, Giannini et al. 2013, Whelan et al. 2014).

HH34~IRS is an embedded low-mass object ($M_* \sim$0.5 M$_\odot$, $L_* \sim$ 15 L$_\odot$,
Antoniucci et al. 2008) located in the L1642 cloud ($d$=414 pc) and driving 
the well-known HH34 parsec scale flow. The flow and its associated HH objects are
 driven by a collimated blue-shifted jet, which extends for about 30\arcsec\,  and
 is bright both in optical and IR images (e.g., Reipurth et al. 2002, Antoniucci et al. 2014a). 
This  makes HH34 an ideal target to exploit
 the X-shooter multiwavelength capabilities.
 
High resolution optical imaging and spectroscopy have  mainly been used to study 
the morphology and kinematics of the jet (e.g., Reipurth et al. 2002; Beck et al. 2007;
Raga et al. 2012). Multi-epoch HST observations have  indicated 
that the jet is subject to
variations in velocity and jet axis direction, which can be due to precession of
the outflow source around a possible companion.
Infrared spectroscopy has revealed for the first time the presence of a 
counter-jet (Garcia Lopez et al. 2008), later imaged with Spitzer (Raga et al. 2011).
Several spectroscopic studies in the infrared have been conducted on the HH34 jet,
showing that it emits copiously  in both \feii\, and H$_2$ transitions
(Takami et al.2006, Garcia Lopez et al. 2008, Davis et al. 2011).
An analysis combining both IR and optical spectroscopy has been
applied by Podio et al. (2006) to infer the jet excitation structure at large scale. 

Taking advantage of the higher sensitivity, resolution, and spectral 
coverage of the X-shooter, we can now derive better estimates on the physical
conditions at the base of the HH34 micro-jet and  investigate  the relationship
between gas excitation and kinematics.  This information can be  used to
constrain the mechanism responsible for the jet formation and to discuss 
the similarities and differences with jets in CTT stars.

The paper is structured as follows. Section 2 describes the X-shooter observations 
and data reduction, while in Sect. 3 we present the main results of the line detection and
kinematics of the HH34 small-scale jet. In Sect. 4 we present
the diagnostic analysis applied to the jet emission, separately
discussing the inner jet components and the variation of excitation
along the jet. 
In Sect. 5 we use the luminosity of selected permitted
lines to put constraints on the source mass accretion rate.
Finally, Sect. 6 summarizes and discusses the main outcome of our analysis.


   \begin{figure*}[t]
\includegraphics[angle=0,width=5cm]{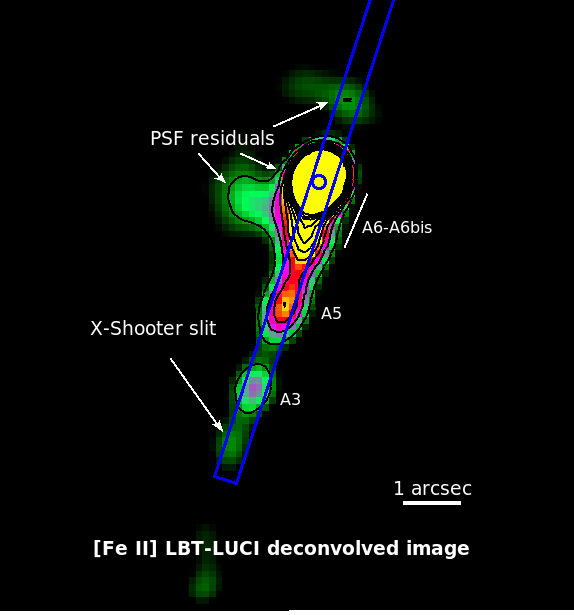}
\includegraphics[angle=0,width=10cm,trim={0 5cm 0 10cm},clip]{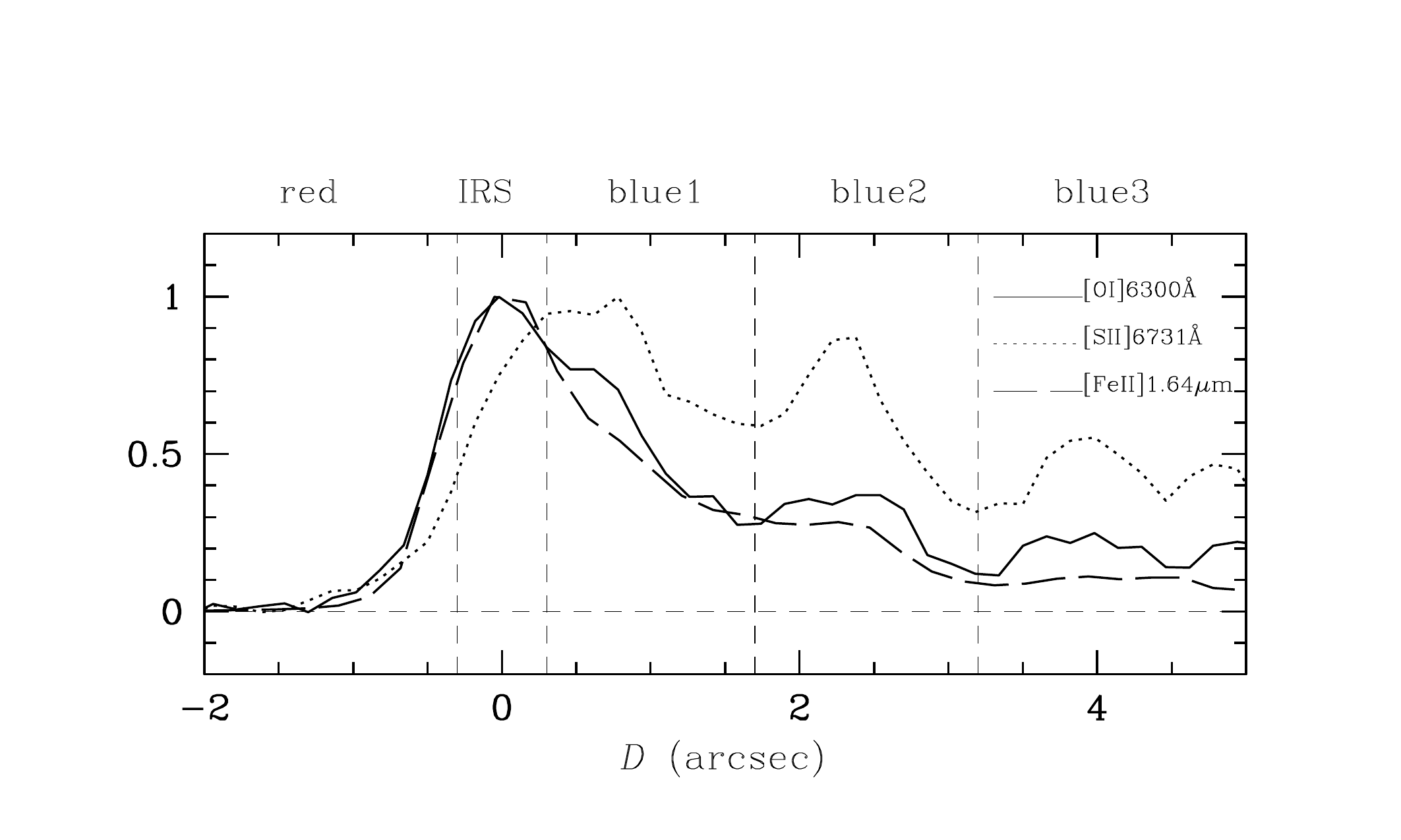}
      \caption{\textit{Left:} Adopted X-shooter slit  over a 
      [Fe II] 1.64$\mu$m narrow band image of the HH34 jet obtained with LBT-LUCI 
      (Antoniucci et al. 2014a). The blue circle indicates the source position at
  $\alpha$ = 05$^h$ 35$^{m}$ 29\fs 9 and $\delta$ = $-$06$\degr$ 26$\arcmin$ 58$\arcsec$.      \textit{Right:} Normalized intensity profiles of \oi\, 6300\AA,
      \sii\,6731\AA,\ and \feii\,1.64\um\ along the HH34 jet. Distance in arcsec is measured
      with respect to the HH34 IRS source. The spatial regions we adopted for the spectral
      extraction are indicated by dashed vertical lines.}
         \label{fig.1}
   \end{figure*}

      \begin{figure*}
\includegraphics[angle=0,width=15cm,trim={0 5cm 0 7cm},clip ]{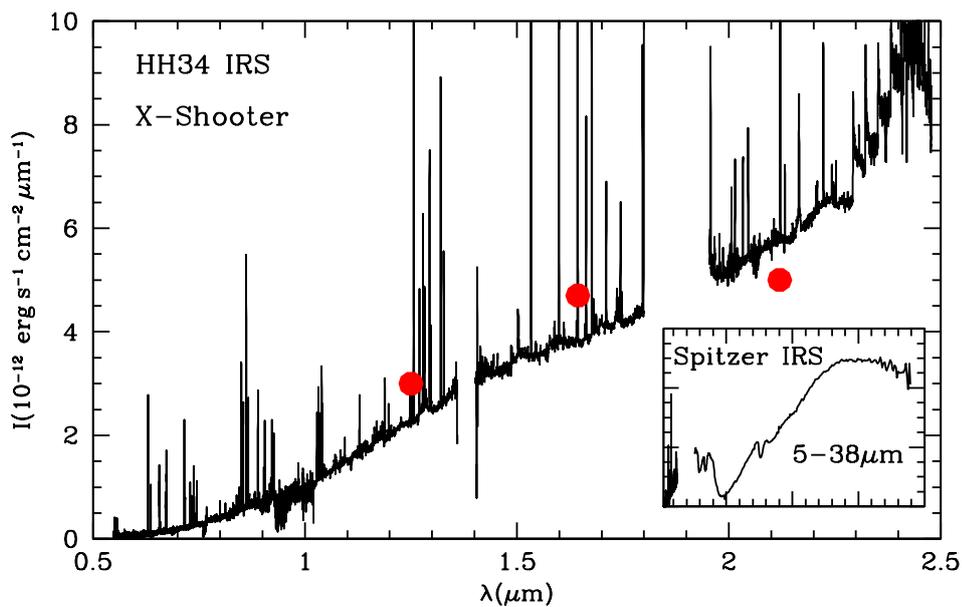}
     \caption{X-shooter spectrum extracted at the HH34~IRS source position. Red dots correspond to the 2MASS photometry in the $J, H$, and $K$ bands. The inset shows the mid-IR archival \textit{Spitzer}-IRS spectrum. 
              }
         \label{fig.2}
   \end{figure*}

   \begin{figure*}[t]
\includegraphics[angle=0, width =15cm]{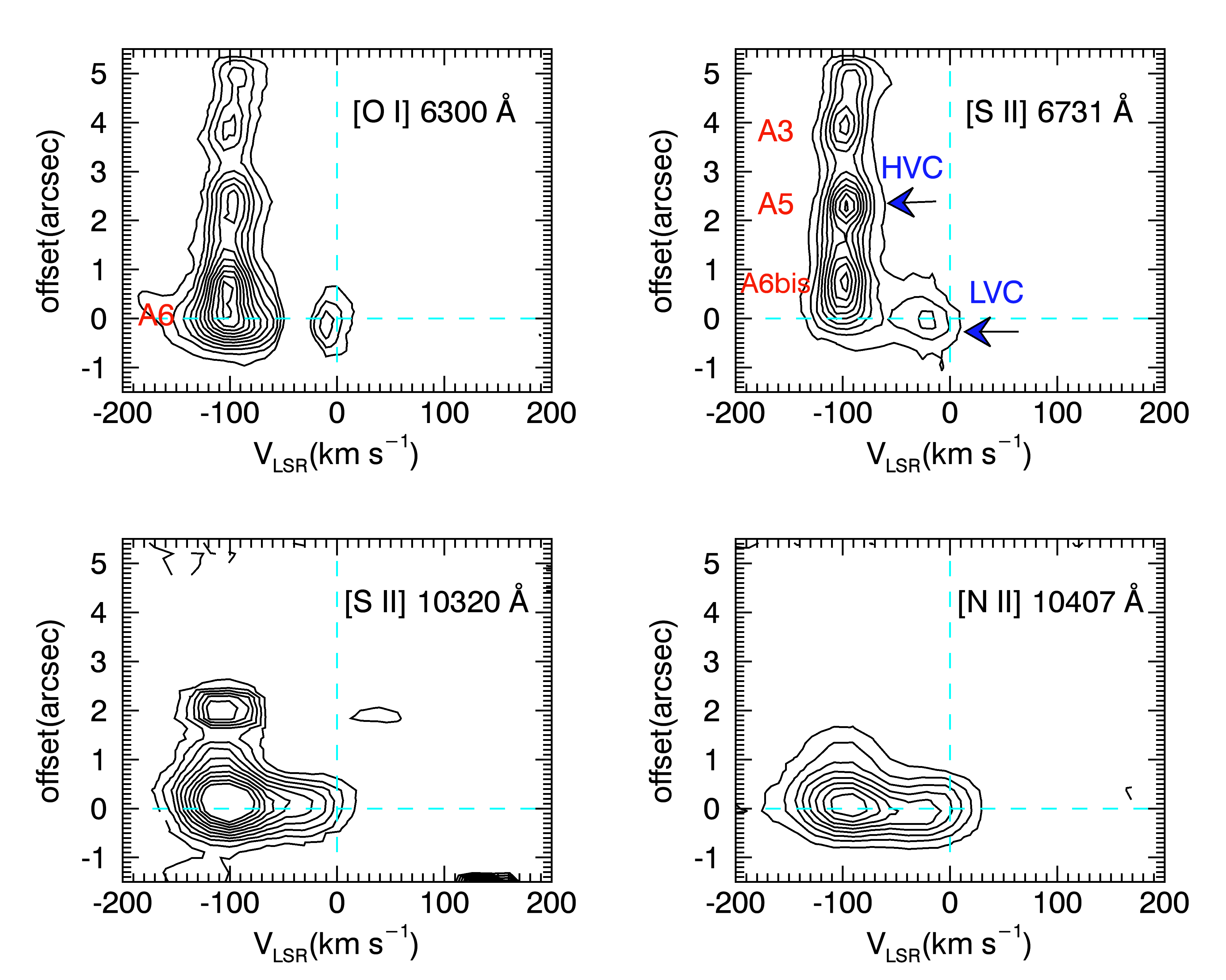}
      \caption{Continuum-subtracted position velocity (PV) maps of \oi , \sii,\, and \nii\, lines observed in the
      X-shooter spectrum. The PV are oriented along the jet, with a PA = 162$\degr$. 
       Contours start at 3$\sigma$ and are drawn at steps of 3$\sigma$ with the exception of \sii\, 6731\AA\, where
       the contours are drawn in steps of 6$\sigma$. Radial velocities have been corrected for the 
       source systemic velocity of $+$8\kms. The main peaks are labeled following the nomenclature
       of Fig. 1. The HVC and LVC are also indicated.
      }
         \label{fig.3}
   \end{figure*}
   
      \begin{figure*}[t]
\includegraphics[angle=0, width =15cm]{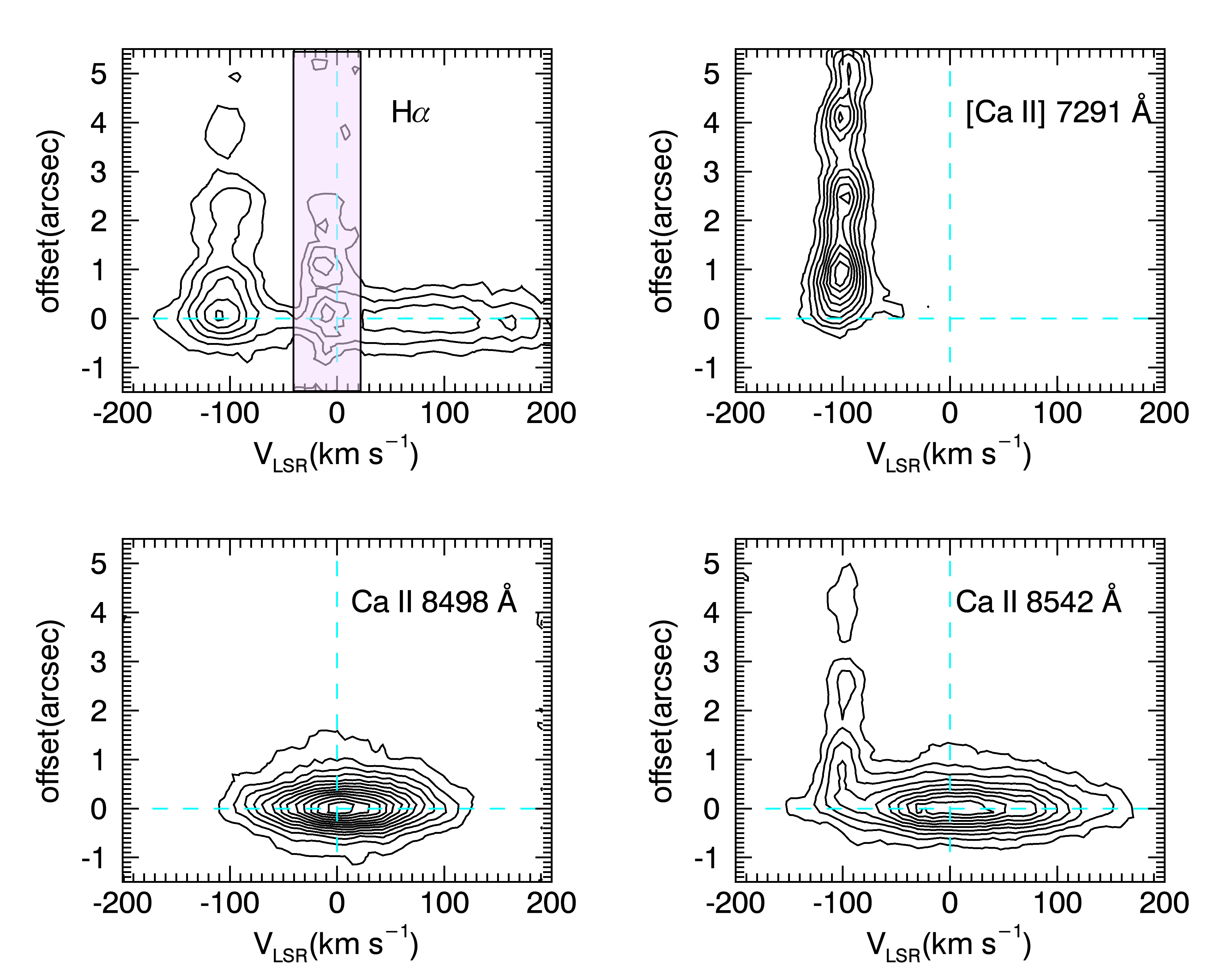}
      \caption{As in Fig. 3 but for H$\alpha$ and \ion{Ca}{II} lines. Contours start at 3$\sigma$ and are drawn at steps of 3$\sigma$.
      The shadowed area in the H$\alpha$ PV indicates a region with strong contamination from telluric residuals 
      }
         \label{fig.4}
   \end{figure*}

   \begin{figure*}[t]
\includegraphics[angle=0, width =15cm]{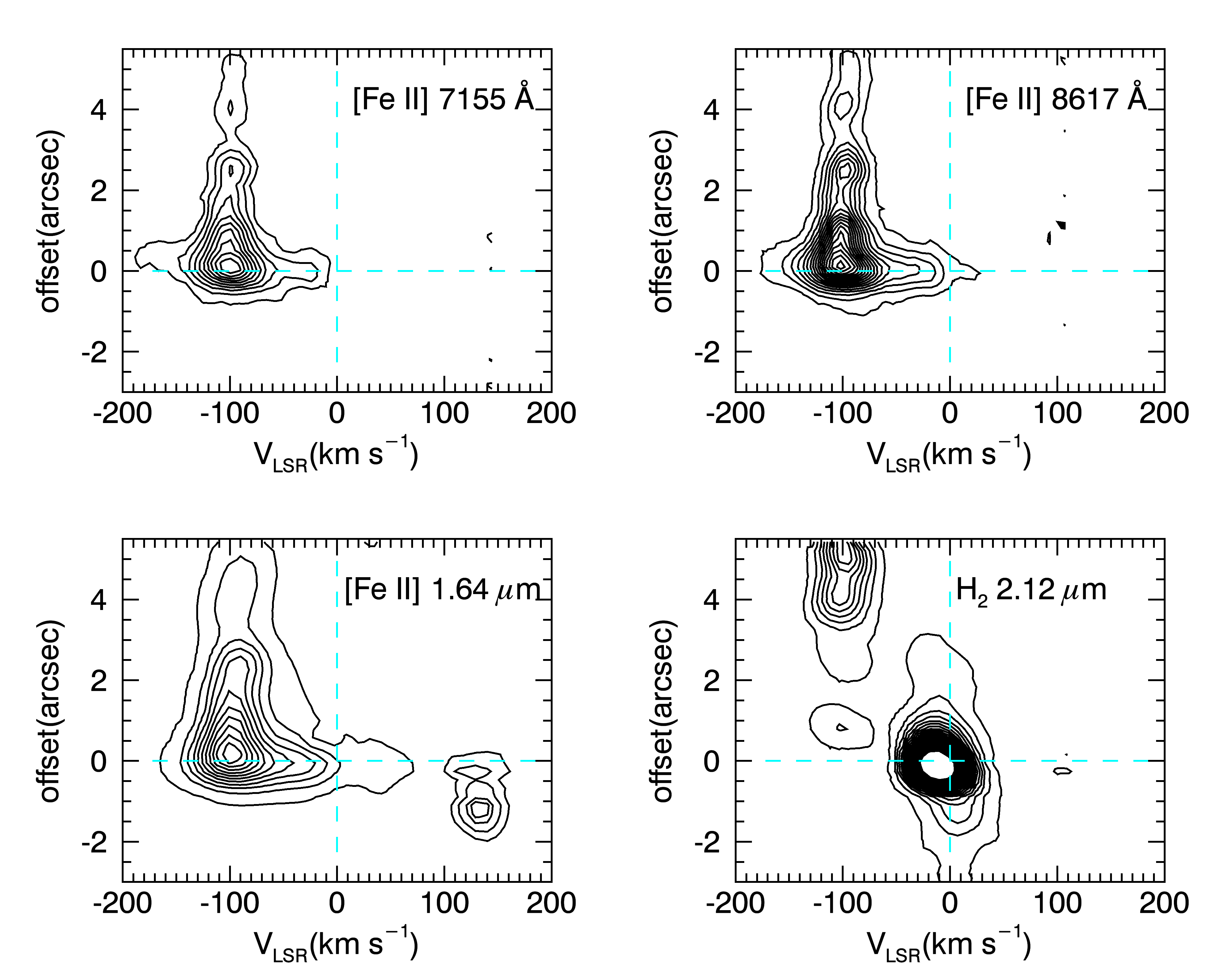}
      \caption{As in Fig. 3 but for \feii\, and the H$_2$ 2.12\um\, lines. Contours start at 3$\sigma$ and are drawn at steps of 3$\sigma$, 
      with the exception of 1.64\um\, where the first five contours are at 2, 4, 6, 8, and 10$\sigma$ to emphasize the red-shifted emission, and then continue at steps of 6$\sigma$.}
         \label{fig.5}
   \end{figure*}

\section{Observations}
X-shooter observations of the HH34~IRS jet were conducted during the 
period November 2012-January 2013. Four different observations were acquired, each
one integrated for 30 min on-source.  
The 11\arcsec\, slit was centered on the HH34~IRS source ($\alpha$ = 05$^h$ 35$^{m}$ 29\fs 9, $\delta$ = $-$06$\degr$ 26$\arcmin$ 58$\arcsec$) 
and aligned with the jet axis at P.A. 162$\degr$ (see Fig. 1). 
The  seeing during the observations was always between 0\farcs6 and 1\farcs0 . 
Since the slit is filled with the emission from the jet, we performed off-source nodding 
for sky subtraction. Slit widths of 0\farcs 5, 0\farcs 4, and 0\farcs 4 were used
for the UVB, VIS, and near-IR (NIR) arms, yielding resolutions of 9900, 18200, and 7780, respectively.
The pixel scale is 0\farcs 16 for the UVB and VIS arms, and 0\farcs 21 for the NIR arms. 
Data reduction was performed using the X-shooter pipeline version 2.2.0 (Modigliani et al. 2010), which provides 2D calibrated spectral images. Post-pipeline processing was performed using
the \textit{IRAF}\footnote{IRAF is distributed by the National Optical Astronomy Observatory, which is operated by the Association of Universities for Research in Astronomy (AURA) under a cooperative agreement with the National Science Foundation.} package.
This includes sky subtraction and telluric correction in the NIR images. 
To this end, the spectra of telluric standard stars acquired at similar airmass as the objects were used after the intrinsic photosperic features were removed. 
Finally, we combined the four different spectral images
applying a median filter to get the final 2D spectral image. 

Flux calibrations were performed within the pipeline using spectro-photometric standard stars
acquired during the same night as the observations. In order to check for inter-calibration
between the different arms, we  extracted a spectrum of the HH34~IRS object
in an aperture of 3\arcsec centered on-source and examined the continuum shape. 
Basically no continuum is detected shortward of 5000\AA, while  there is a perfect match
between the VIS and NIR arms in their overlapping region. However, comparing the NIR
fluxes with the 2MASS photometry (Skrutskie et al. 2006) of the source we noticed that the X-shooter spectrum 
underestimates the NIR photometric points, likely because of  
slit losses due to the narrow slit width of the observations. 
A correct evaluation of the slit-losses should take into account the geometry
of the continuum emission, as the source cannot be considered point-like owing
to the presence of a diffuse nebulosity. 
We have instead re-scaled the X-shooter 
spectrum by a factor that minimizes the difference between the X-shooter fluxes and
the 2MASS photometric points.
The applied correction amounts to a factor of 2.7. 
However, as can be seen in Fig. 2, the shape of the 2MASS continuum is not  reproduced well.
 This could be due to the presence of the diffuse nebulosity enhancing 
 the $J$ and $H$ bands in the 2MASS large apertures, but also to source intrinsic variability. 

Given the above discussion, we estimate that the relative flux calibration error between  
the three arms is very low, not more than 10\%, while the absolute
flux calibration is more uncertain and can be as high as 50\%.

\section{Results}
As shown in Fig. 1, where the X-shooter slit is drawn over a \feii\, 1.64\um\, image
of the HH34 inner jet, our observations cover several emission peaks belonging to the
blue-shifted jet (Antoniucci et al. 2014a). 
Separate spectra have been extracted from the
2D spectral images, corresponding to different regions selected from the intensity 
profiles of bright lines (see Fig.1, right panel) -- 
IRS, blue1, blue2, and blue3 -- at progressively larger distances from the central object.
These regions roughly comprise the knots called A3-A6 in previous studies (e.g., Garcia Lopez et
al. 2008); however, given  the large proper motion of these knots (see, e.g., Antoniucci et al. 2014a
and Section 3.1), we prefer to use a different nomenclature for our extraction regions 
that does not necessarily correspond to previously observed emission peaks.

The red-shifted counter-jet is clearly seen  
in the  large-scale \feii\, image of Antoniucci et al (2014a). In Fig. 1 counter-jet
emission is contaminated by the residuals of the PSF subtraction. However, 
in Garcia-Lopez et al. (2008) red-shifted emission is detected down to the central source.
We therefore extract an additional spectrum, named red, comprising 
the counter-jet region close to the source. 

Figure 2 shows the HH34~IRS complete spectrum. As expected for an embedded class I source, the spectrum sharply rises from UV to the infrared. 
Continuum and line emission are detected only above $\sim$5000\AA\,. The X-shooter spectrum
is also compared with a \textit{Spitzer}-IRS spectrum taken from the \textit{Spitzer}
Heritage Archive (Program ORION$\_$IRS, P.I. T. Megeath).

Strong emission lines are observed in the HH34~IRS spectrum, which are reported in Table A1-A3 of the Appendix, together with their identification and fluxes\footnote{Transition wavelengths in Table 1 are air wavelengths expressed in Angstrom. In the paper, we  refer for convenience to lines in Angstrom or in \um\, depending on whether they are below or above 1 micron.}. Lines detected in the other apertures are reported in Table A4.  Forbidden lines from abundant 
species originating in the jet are detected up to excitation energies of $\sim$ 31\,000 cm$^{-1}$. 
In addition, we identify permitted lines from hydrogen (from the Brackett and Paschen series), helium, \ion{O}{i}, \ion{Ca}{ii}, and \ion{Na}{ii}. 
Additional  features, especially in the IR arm, do not have a clear identification, but the line shape and velocity widths suggest they are associated with metallic lines emitted in the same region as the other identified permitted lines. 
Finally, molecular lines from H$_2$ and CO ro-vibrational transitions are also detected. 
The spectra extracted along the jet are, as expected, dominated by forbidden lines; however, H$\alpha$ and \ion{Ca}{ii}~8542\AA\, are also detected far from the central source. 
This is shown in Figs. 3, 4, and 5, where  the position-velocity (PV) diagrams of a few representative forbidden, permitted, and molecular transitions are presented.
The local continuum has been removed by performing a linear fit through the emission adjacent to the lines. Since no photospheric lines of the source are detected -- owing to the large veiling that fills all the absorption features -- the source systemic velocity has been corrected for the velocity of the parental cloud (+8\kms , Anglada et al. 1995) after having corrected the spectra for the relative motion of Earth and Sun with respect to the standard of rest with the \textit{IRAF} task \texttt{rvcorrect}.

Forbidden lines (e.g., \oii, \sii, \feii, [\ion{N}{i}])
have an extended blue-shifted component peaking at $\sim$ $-$100 \kms\ (hereafter high velocity, HV, or 
high velocity component, HVC) related to the jet. 
In addition, a spatially unresolved emission at lower velocity is also detected in all the lines
(low velocity, LV, or low velocity component, LVC). 

Finally a red-shifted HVC (V$_r \sim$+130\kms) is clearly seen in the \feii\,1.64\um\, line. 
This component is not detected at shorter wavelengths and it is only barely detected in a few weaker
\feii\, lines longward of 1.64\um .  We think this is due to the large extinction of the counter-jet. 
The non-detection of the 1.25\um\, line set a lower limit on the extinction in the red-shifted
component to $A_V >$ 13 mag, assuming the intrinsic 1.25\um/1.64\um\, ratio given in Giannini et al. (2015b) (see discussion in Sect. 4.1.2).

 The above kinematical signatures were already known from previous \feii\, IR observations of 
 HH34 (Garcia-Lopez et al. 2008, Antoniucci et al. 2008). 
 We can now discuss any differences in the kinematical properties of lines having different excitation conditions. 

To this end, we  performed a Gaussian fitting to the line profile,
 from which the line V$_{peak}$ and velocity width were estimated.
 For the on-source spectrum, parameters for the HVC and LVC components were
 separately measured by fitting the line profile with two Gaussian 
 components. Fluxes for all the lines derived from this Gaussian fitting are listed in
  Tables A1-A4, while Table 1 gives the kinematical parameters 
  of a subsample of lines. 
 
From Table 1 and from the PV diagrams  no shifts 
in the peak velocity of the HVC are appreciated within the errors, 
since V$_{r}$ is $\sim$ ($-$100$\pm$10)\kms\, 
for all the different lines, irrespective of their ionization stage or excitation temperature.

The jet velocity also remains fairly constant  in all the extracted positions.
Regarding the LVC, we observe that the peak velocity of \feii\, and \caii\,  lines
($\sim$ 40-50\kms) is higher than for \sii\, and \oi\,  lines ($\sim$ 25-30\kms).
This difference could be related to a large depletion of refractory elements at the
lowest velocities. We  discuss this aspect in Section 4.3.

No variations of the line widths are appreciable at the X-shooter resolution.
However, we do  see a decrease in $\Delta$V values from the inner ($\Delta$V $\sim$ 50$\pm$10\kms\, on-source) to the outer ($\Delta V$ $\sim$ 35$\pm$10\kms\, in blue2/3) knots.

Comparing the different PV diagrams, we note that high excitation lines 
(such as \nii\, 10407\AA\, and \sii\,10320\AA, with $T_{ex} \sim$ 35\,000-40\,000 K) are spatially more compact than the other lines, as they are only detected  within 2\arcsec\, of the source. 
The first emission peak is, for the majority of the lines, shifted by $\sim$0\farcs 3 
from the source position. Exceptions are \sii\,6731\AA\, and [\ion{C}{ii}]7291\AA, which have the first peak at about 0\farcs 7. As discussed in the next section, this is probably due to the low critical density of these lines, which are therefore quenched in the high density region closer to the star. Such a secondary peak is also seen in 
the \oi\, line. At variance with the HVC, the LVC always peaks  on-source.

Permitted lines, such as those of \ion{Ca}{ii}, peak on-source at V$_r$ $\sim$ 0 \kms . However, in the bright  \ion{Ca}{ii} line at 8542\AA\, the extended emission from the jet is also clearly detected, showing the same morphology and kinematics 
as the \caii\, 7291\AA\, emission. 
Such a jet emission is also detected  in the \ion{Ca}{ii} 8662\AA\, but not in
 the \ion{Ca}{ii} 8498\AA\,line, probably owing to sensitivity. The
H$\alpha$ line presents a compact and very broad emission, which is largely self-absorbed at zero velocity, 
and  an extended HV jet emission. 
Finally, H$_2$ has kinematical signatures different from
the atomic lines. 
The HVC component is prominent only at large distances from the central source ($\ga$ 4\arcsec), 
while the LVC is resolved and extends up to
$\pm$ 2\arcsec\ from the source. The H$_2$ kinematics of HH34 has   been
presented and discussed in Garcia-Lopez et al. (2008) and is not  further addressed here.

\setlength{\tabcolsep}{3pt}
\begin{table*}
\caption[]{Peak velocities and width of selected lines}
\vspace{0.5cm}
\begin{tabular}{c|cc|cc|cc|cc|cc|cc|cc|cc}
\hline
& \multicolumn{2}{c}{[FeII]8617} & \multicolumn{2}{c}{[FeII]12566} & \multicolumn{2}{c}{[FeII]16435} & 
\multicolumn{2}{c}{[SII]6730} & \multicolumn{2}{c}{[SII]10336} & \multicolumn{2}{c}{[OI]6300} & 
\multicolumn{2}{c}{[NII]6583} & \multicolumn{2}{c}{[CaII]7291}\\
& V$_r$ & $\Delta$V & V$_r$ & $\Delta$V & V$_r$ & $\Delta$V & V$_r$ & $\Delta$V &V$_r$ & $\Delta$V &V$_r$ 
& $\Delta$V &V$_r$ & $\Delta$V & V$_r$ & $\Delta$V \\
\hline
IRS HV & -104 & 48& -109 & 48& -107 & 47& -104 & 53& -106& 35 & -104& 59& -102& 39& -106& 36\\
IRS LV & -42 & 59& -51 & 64 & -52  &  66 & -32 & 58 & -31 &52 & -25& 43& ...&&  -47& 49\\
\hline\\
blue 1& -103& 31 &-107& 36&-105& 36&-103& 30&-104& 43 &-105& 29& -109& 43 &-102& 32\\
blue 2& -100& 26 &-101& 33& -110& 34 &-75& 27&-101& 43 &-102& 29& -104& 43&-99& 25\\
blue 3& -99 & 26 &-101& 31 &-110& 33 &-99& 26 &...& ... &-102& 29&...& ...&-99& 25\\ 
\hline\\[-5pt]
\end{tabular}
\\
\footnotesize{Notes:V$_r$ and $\Delta$V (in km\,s$^{-1}$) are the peak radial velocity and FWHM of the Gaussian fit through the line profile. The peak velocity is measured with respect to the cloud velocity of 8\kms . Errors on velocity peaks are on the order of 5\kms .  }
\end{table*}


\subsection{Proper motion analysis}

Observations of the HH34 jet over several  decades have shown that the jet has a 
significant proper motion, with emission knots that change in position
and morphology and new peaks that appear on timescales of a few years (Hartigan et al. 2011; Raga et al. 2012; Antoniucci et al. 2014a).
Our PV data, compared with observations taken in different periods, can be used to 
measure the proper motions (PMs) of the knots close to the source.
To this aim, we compare our data with those presented in
 Garcia Lopez et al. (2008), consisting of \feii\, spectra acquired with the ISAAC instrument
(spatial sampling 0\farcs 146/pixel, thus comparable to that of the X-shooter VIS arm)
in December 2004, i.e., 8 years prior to our observations. We also used  data with adequate 
spatial resolution, namely those of Davis et al. (2011) (SINFONI \feii\, data from 
Oct. 2006, 0\farcs 05/pixel) and Antoniucci et al. (2014a) (LBT \feii\, narrow band imaging, data from Apr. 2013, 0\farcs 118/pixel). In the latter data, only the region outside  $\sim$ 1\arcsec\, can be used
owing to the contamination by residuals of the central source continuum deconvolution  (see Fig.1). 
Table 2 summarizes the shifts of the main emission components with respect to the central source 
position identified through the continuum fit.

The shifts on the X-shooter spectrum were measured from several bright lines,
namely \oi\, 6300\AA, \sii\, 6731\AA, \caii\, 7291\AA, \feii\, 8617\AA, and \feii\, 1.64$\mu$m,
and the results were averaged.
In particular, we clearly identify three emission knots that can be associated with the
knots A6, A5, and A3 of the Garcia Lopez et al. paper (see also Figs. 1 and 3). In addition,
and as described before, the \sii\, 6731\AA\ and \caii\, 7291\AA\  lines present an emission
peak at 0\farcs 7 that was not identified in previous ISAAC or SINFONI \feii\, observations; 
 here we name it  knot A6bis (indicated in Fig. 3).
From Table 2, we see that knots A5 and A3 have moved during time: assuming
the distance of 414 pc, the displacement corresponds to a tangential velocity 
of 270 and 200\kms\, for the A5 and A3 knots, respectively. The PM of the HH34 
jet has been  measured by different authors. In particular Raga et al. (2012)
compared different epoch HST observations deriving for the A5 and A3 HH34 jet knots
a V$_t$ comparable to our derived values (namely 250 and 210\kms). 

We  find, however,  that inner knot A6 has not appreciably changed
its position. The derived shift is compatible with the 
0\farcs 2 shift measured by Garcia Lopez within the error, corresponding to an
upper limit on the tangential velocity of $<$37\kms. Davis et al. (2011) measured a slightly
lower offset, but their value is still compatible with the above upper limit.

 The same A6 peak is present in observations taken in three different dates, which
makes it very unlikely that we are seeing different ejections that 
are located by chance at the same offset during the observations.
We can also exclude that the emission that we see is actually located on-source, but
observed with an offset due to the large on-source extinction; in this case we would see
a wavelength dependence of the offset while all the considered line peaks
at the same position. 
The presence of stationary emission knots in the inner jet region is actually not
a peculiar feature, as it has been already observed in other jets at different 
wavelengths, i.e., in DG Tau (offset 35-50 AU from source; Schneider et al. 2013, 
White et al. 2014), L1551-IRS5 (X-ray stationary emission at
$\sim$ 70 AU; Bonito et al. 2011), SVS13 (\feii\,1.64\um, emission at 20 AU from source;  
Davis et al. 2008).
Such stationary structures are usually interpreted, in  
magneto-centrifugally driven jets, as shocks occurring at the jet recollimation
region. In this framework, as the flow is launched outward,  it enters  a region where 
the magnetic tension toward the jet axis exceeds the centrifugal
force so that the jet begins to recollimate at super-Alfv\'enic speeds, thus producing a shock region. In HH34 the stationary shock occurs at $\sim$ 80-100 AU, thus at a distance larger 
than those observed in the other jets, but still consistent with models
that predict that the recollimation shock region should be located at tens of AU 
above the circumstellar disk (e.g., Gomez de Castro \& Pudritz, 1993).

\begin{table*}
\caption[]{Offsets of emission peaks and tangential velocity}
\vspace{0.5cm}
\begin{tabular}{cccccc}
\hline
       & Garcia Lopez et al. (2008)$^a$& Davis et al. (2011) &This work$^b$ & Antoniucci et al. (2014a)$^c$ & <V$_t$>\\
       & Dec. 2004   & Oct. 2006           & Nov. 2012  & Apr. 2013 & \\
       &                   &     &     & \kms\\
\hline
A6     & 0\farcs 2     &   0\farcs 093$\pm$ 0\farcs 007     & 0\farcs 25$\pm$ 0\farcs 05 & ... & $<$37\\
A6bis  & ...           &   ...            & 0\farcs 65$\pm$ 0\farcs 05$^d$ & ... & ...\\
A5     & 1\farcs 2     &   0\farcs 91$\pm$ 0\farcs 07                & 2\farcs 2$\pm$ 0\farcs 1 & 2\farcs 4$\pm$ 0\farcs 1& 270$\pm$20\\
A3     & 3\farcs 1     &   ...               & 4\farcs 0$\pm$ 0\farcs 5 & 3\farcs 9$\pm$ 0\farcs 1& 207$\pm$20\\
\hline\\[-5pt]
\end{tabular}
\\
$^a$ from \feii 1.64\um\, long-slit spectra. \\
$^b$ measured as an average of the shifts in the lines \oi 6300\AA, \sii 6731\AA\ (except for knot A6), \caii 7291\AA,
\feii 8617\AA, \feii 1.64\um\\
$^c$ from the \feii 1.64\um\, image
\end{table*}


\section{Jet physical conditions }

The wealth of forbidden lines detected in the HH34 jet allows  a detailed
line diagnostic analysis to be performed on the gas physical conditions.
Previous studies have already addressed the excitation conditions in the HH34 jet
employing different set of diagnostic lines. 
Podio et al. (2006) considered several optical and IR lines to measure the electron density ($n_e$),
kinetic temperature ($T_e$), and ionization fraction ($x_e$) along the jet. Owing to the limited
spatial resolution and sensitivity, the jet section below $\sim$ 2\arcsec\, was not
sampled and only large-scale variations of physical conditions were derived. 
Garcia Lopez et al. (2006) and Davis et al. (2011) used \feii\, IR lines to infer the 
electron density ($n_e$)  and extinction, and their variations with velocity, in the innermost jet knots.  
A more accurate analysis of the HH34 micro-jet is now possible by adopting lines of 
different species and excitation.

In the next section, we  describe in detail the diagnostic analysis performed
on the spectrum extracted at the source position, which is the richest in lines.
This analysis allows us to investigate the properties
in the initial $\sim$0\farcs 5 section ($\sim$200 AU) where the jet shows 
a more complex kinematical behavior. In particular, we investigate 
any differences among the physical properties of the low and high velocity components.
Variations of physical conditions in the other knots along the jet will be discussed
in a separate section.

\subsection{HH34-IRS jet base}

\subsubsection{Excitation trends as a function of velocity}

\begin{figure*}
      
\includegraphics[angle=0,width=18cm,trim={0 6cm 0 13cm},clip ]{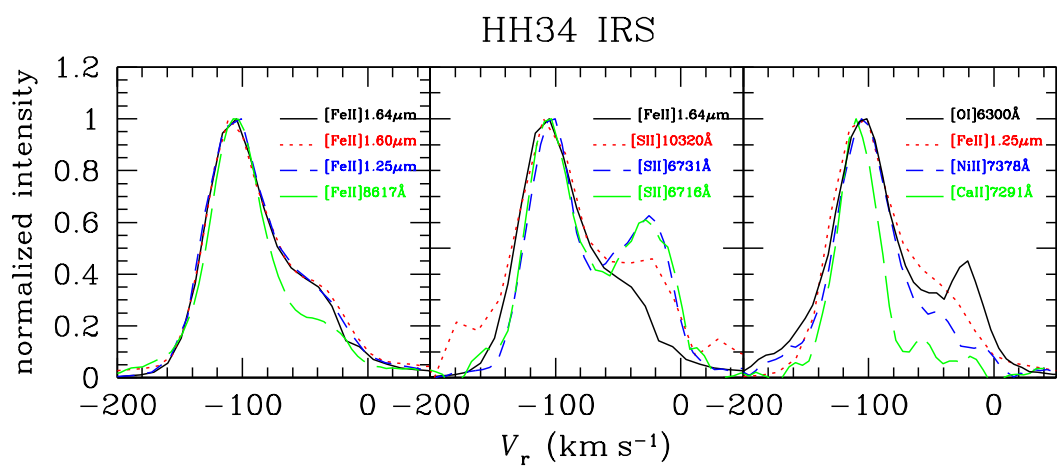}
      \caption{Comparison of continuum subtracted line profiles extracted from the HH34~IRS
         spectrum. Line profiles are normalized to the peak of the HVC.}
         \label{fig.6}
   \end{figure*}

Figure \ref{fig.6} shows the profiles of different forbidden lines extracted from the 
spectrum of the IRS position. Their comparison can be used to infer 
general trends on the difference in excitation between the high velocity jet and the 
component at lower velocity. 

The left panel of Fig. \ref{fig.6} shows the profiles of four 
\feii\, lines that are sensitive to $T_e$, $n_e$, and $A_{\rm v}$ variations 
(e.g., Nisini et al. 2005).
In particular, the ratios \feii 1.64\um/1.25\um\, and \feii 1.64\um/1.60\um\, 
trace extinction and density variations, respectively. The normalized profiles
of these three lines are remarkably similar and this agreement does not
change with velocity, indicating
that electron density and extinction remain constant in the 
HV and LV gas. On the other hand, the 1.64\um/ 8617\AA\ line ratio 
decreases with increasing $T_e$ for a fixed density value: consequently, the
observed profile variations with velocity suggests that 
the LVC is colder than the HVC. 

The above trends are confirmed by the inspection of diagnostic lines from other species. 
The central panel of Fig. \ref{fig.6} 
shows the \feii 1.64\um\, line in comparison with different \sii\, lines. 
The \sii 6716/6731\AA\, ratio, sensitive to $n_e$, does not vary with velocity among the 
LVC and HVC, while the \sii 10320/6731\AA, sensitive to $T_e$, shows a decrease 
in the LV gas. 

From inspection of Fig.6 we also note that 
the \feii/\sii\ (central panel) and the \feii/\oi\ (right panel) ratios decrease from the HVC to the LVC, 
irrespective of the considered \sii\, line. We interpret this as being due to a difference in the gas-phase 
Fe abundance in the two components, suggesting that Fe is more depleted on dust grains 
in the LVC emitting region than in  the HVC region. The same behavior is observed 
in lines from other refractory species, like Ni and Ca (right panel of Fig. \ref{fig.6}). 
This is discussed  more quantitatively  in section 4.3.

\subsubsection{Diagnostics analysis}

The above qualitative considerations can be quantified by comparing the observed line fluxes,
 corrected for reddening, to statistical equilibrium NLTE codes under  the assumption 
that the observed lines are optically thin. To this end, we made use of the 
NEBULIO database (Giannini et al. 2015a; see http://www.oa-roma.inaf.it/irgroup/line$\_$grids/Atomic$\_$line$\_$grids/Home.html)

We first consider the large number of \feii\, lines with different excitation conditions
that can be used to derive the relevant parameters without any assumptions 
regarding elemental abundance. Giannini et al. (2013), have shown that owing to the uncertainty on the 
atomic parameters of the \feii\, lines, the best approach is statistical and implies  
including  the largest number of line ratios in the analysis. 

To this end, we  fitted simultaneously all the lines detected with a
S/N higher than 5. The fit was  performed separately for the HVC and LVC,
whose relative fluxes -- derived from a two-Gaussian fitting of the line profiles --
are listed in Table A1. 
In heavily embedded sources like HH34~IRS, it is of paramount importance to have a good 
determination of the extinction of the line emitting region, especially when ratios of lines
in a wide wavelength range are considered. Extinction can be efficiently derived from
the ratios of transitions sharing the same upper level so as to remove the dependence on the physical
parameters influencing the level populations. Several \feii\, lines can be used for this 
and the 1.25\um/1.64\um\ intensity ratio involves the brightest ones. 
However, instead of using the theoretical radiative rate, which is affected by
a large uncertainty (Bautista et al. 2013), we adopt the empirical determination of the radiative 
rate coefficients given in Giannini et al. (2015b). 
According to this work, the intrinsic \feii\, 1.25\um/1.64\um\, intensity ratio
 is between 1.11 and 1.2, from which
we derive an on-source $A_V$   of 9.0$\pm$0.5 mag, with no variations between the HVC and LVC. 

We have consequently dereddened all the observed \feii\, lines assuming $A_V$ = 9 mag and adopting the extinction law of Cardelli et al. (1989). We have then compared these values against a grid of models in which the density varies from 10$^3$ to 10$^8$ \cmt\, and the temperature from 3000 to 40\,000 K. 

Figure 7 shows the comparison between the observed intensities, normalized to the intensity of 
the 8616 \AA\, line, with the theoretical ratios of the best-fit models for both the HVC and LVC.
The best-fit models are $n_e \sim$5\,10$^4$ \cmt , $T_e >$ 15\,000 K for the HVC and 
 $n_e \sim$5\,10$^4$ \cmt , $T_e \sim$ 7\,000 K for the LVC.
For the HVC, Fig. 7 shows that the model is not able
to reproduce the lines with upper energy $ E_{up} >$ 15\,000 \cmu.
In fact, ratios of lines with high upper energy are 
systematically underestimated by up to an order of magnitude by the single-component model.
These lines, originating from levels above the a$^4\!P$ term,
have very high critical density (e.g., $n_{cr} \sim$ 10$^7$\cmt\, at $T_e$=10\,000 K ). 
Their excess emission could therefore be explained by the presence 
of an additional gas component at higher density.
To test this hypothesis, we  fitted  the lines with $E_{up} > $ 15\,000 \cmu\ separately. 
The considered ratios, involving both UV and IR lines, are very sensitive to extinction variations. 
Conversely, they are rather insensitive to temperature variations
since they all have a similar excitation energy. In Fig. 8 we show $\chi{^2}$-contours obtained 
varying both density  and $A_V$ values: within 90\% of confidence $A_V$ varies 
between 8 and 10, while the density is larger than 10$^6$ \cmt .

The relative contribution of the low and high density gas component in terms of emitting volume can be
evaluated by comparing the absolute and dereddened observed fluxes with the emissivity values given by the
model. We  assume here that the emission of the 1.25\um\, line is dominated by the low
density gas, while the emission of the 5158\AA\, is dominated by the high density gas.

The ratio between the two emitting volumes -- low density (LD) vs high density (HD) components) -- 
 is $\sim$ 5000, for $n_e$(HD) $\sim 5\,10^6$\cmt . Assuming 
a spherical emitting volume, the typical length scales are $\sim$ 60 AU (0\farcs 15) for the LD component and $\sim$ 3 AU (0\farcs 001) for the HD component.  
These values have at least a factor of two of uncertainty, considering both the 
uncertainty on calibration and extinction. 
Figure 7 shows the comparison between the observed and modeled ratios
obtained by summing  these two components with their relative filling factors. 

In conclusion, the \feii\, analysis suggests the presence of
 an extended LD component with $n_e \sim$5\,10$^4$ \cmt\,
 and spatial scale $\sim$ 60 AU (comparable to the 
length of the slit) and an unresolved (spatial scale 3AU) HD component 
with $n_e \ga$ 10$^6$ \cmt\,. 
The ratio between the 5158\AA\, line observed on-source and in the blue~1 extracted spectrum is $\sim$ 14, while the 8616\AA\ line has a comparable brightness in the two positions, which proves that the HD component represents a compact emission localized close to the source. 
The same is true for the 2.046\um\, line, which is a factor of 10 brighter on-source
with respect to the blue 1 position. The spatial resolution of our observations
does not provide direct clues about the origin of this HD component.
One possibility is that it traces a strong density gradient connected with the collimation shock region discussed in section 3.1. 
Table 3 summarizes all the discussed 
physical parameters inferred from the Fe analysis.

\begin{figure}
      
\includegraphics[angle=0,, width = 9cm,trim={0 5.8cm 0 5.8cm},clip ]{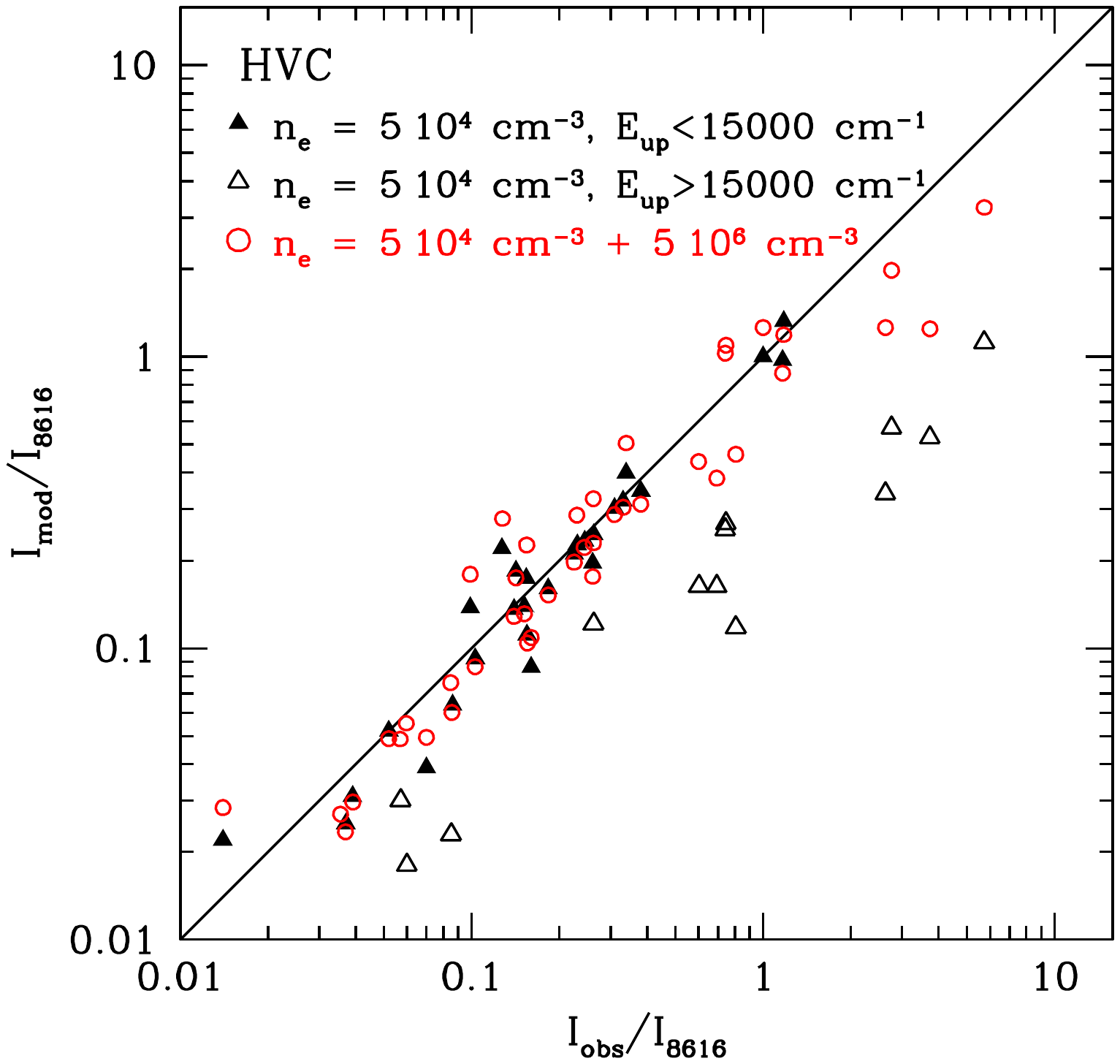}
\includegraphics[angle=0,, width = 9cm,trim={0 5.8cm 0 5.8cm},clip ]{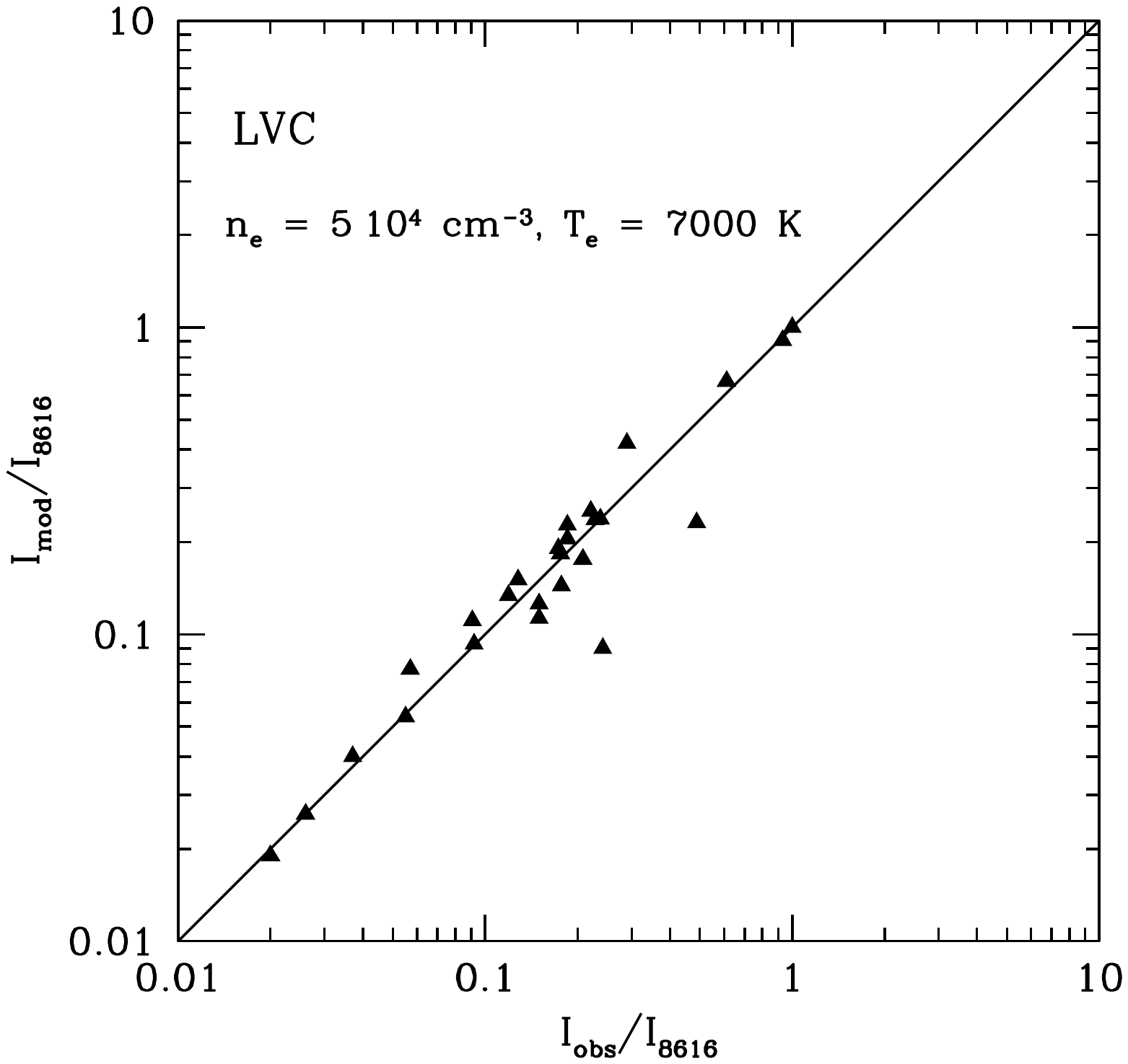}
      \caption{[Fe II] observed vs model predicted line intensities, normalized to the 8616\AA\,
      transition, relative to the HVC (top panel) and LVC (bottom panel). 
      For the HVC, triangles refer to the model with a single gas component at 
      $n_e$ = 5\,10$^4$~\cmt and $T_e > 15\,000$ K:
      filled (open) symbols indicate transitions with upper energies below (above) 15\,000 \cmu. 
      Red open circles refer to the model that includes an additional high density
      component with $n_e >$ 5\,10$^6$~\cmt\ to better fit the transitions at higher excitation
      energies.
              }
         \label{fig.7}
   \end{figure}

\begin{figure}
      
\includegraphics[angle=0, width = 8cm, trim={0 0 0 7cm}, clip]{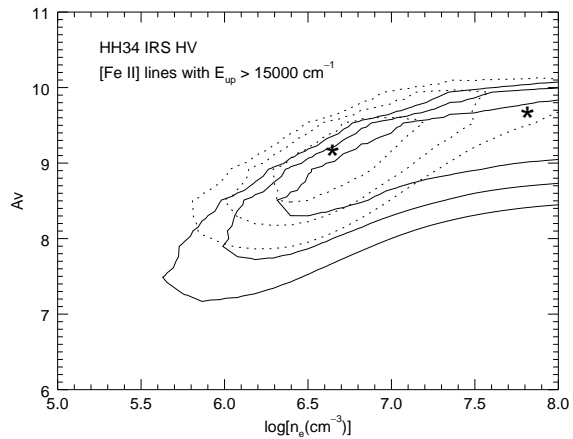}
      \caption{$\chi^2$  contours of the fit through the \feii\, lines observed
      in the HVC and having $E_{up} >$ 15\,000 \cmu. Solid and dotted curves refer to
      $T_e$ = 10\,000 K and 40\,000 K, respectively. The curves refer to increasing values
      of  $\chi^2$ of 30\%, 60\%, and 90\%. The minimum $\chi^2$ value is indicated with a starred
      symbol.
              }
         \label{fig.8}
   \end{figure}   

In addition to Fe, other forbidden lines can be used to derive independent
estimates of the physical conditions. We combined, in particular, \sii\, transitions
at 6716, 6731, and 1.03\um\, to infer $n_e$ and $T_e$ (see Fig. 9). 

Table  3 includes the results of all the above diagnostic analysis.
Temperatures inferred from \sii\, for the HVC and LVC are in agreement 
with the values derived from the \feii\, analysis, 
while the values of the density are slighter lower. This is indeed a 
general result in stellar jets (e.g., 
Giannini et al. 2013; Podio et al. 2006; Nisini et al. 2005) and it is 
due to the low critical density of the \sii\, lines making them more
sensitive to low density regions when density gradients are present. 

Constraints on the ionization fraction can be also obtained 
from the upper limit on the \oii\,7329\AA\, and \feiii\, 5270\AA\, lines,
and from the [\ion{N}{i}]\,5198/\nii\,6583\AA\, detected ratio.
Adopting the ionization equilibrium model described in Giannini et al. (2015a), 
only upper limits on the ionization fraction can be inferred, and all the methods
indicate a low $x_e$ value (see Table 3).
The most stringent upper limit is found from the 
[\ion{N}{i}]/\nii\  ratio which is consistent with $x_e <$ 0.1. 

From the $x_e$ upper limit we derive a lower limit on the total density n$_H >$ 5\,10 $^{5}$\cmt\,
in the HVC. This value can be used to infer the jet mass loss rate by directly
measuring the mass flowing through a jet annulus, namely  \.M$_{jet}$ = $\mu\,m_H\,n_H\times\pi\,r_J^2\,V_J$. Here, $\mu$ = 1.24 is the average atomic weight, $r_J$ the jet radius, and $V_J$
the jet velocity. This method assumes that the jet section is 
completely filled with gas at the considered total density. The value of  
$V_J$ is taken equal to 290\kms, as derived from combining the jet radial and tangential
velocities discussed in section 3. The value of  $r_J$ is assumed equal to the width of the \feii\, emission
as measured by Davis et al. (2011), namely $\sim$ 0\farcs1 ($\sim$40 AU). 
With  a total density of 5\,10$^{5}$\cmt,  \.M$_{jet}$ =  5\,10$^{-7}$ M$_{\odot}$\,yr$^{-1}$  is derived. 
This value is comparable with that derived by Podio et al. (2006) at a few arcsec distance 
from the source adopting the same method, while it is higher by an order of magnitude 
than the value estimated by Garcia Lopez et al. (2008) on the same spatial scale
from the luminosity of the \feii\, lines. This  method, however, is subject to 
more uncertainties as it strongly depends on the adopted extinction value, on
flux losses within the slit, and on the exact iron depletion factor in the jet.  
The derived mass flux rate implies that the source mass accretion rate should be at 
least  $>$ 10 $^{-6}$ M$_{\odot}$\,yr$^{-1}$
to find consistency with any of the jet formation models predicting \.M$_{jet}$/\.M$_{acc} <$0.5
(Ferreira et al. 2006).  This point is further discussed in Section 5.

\subsection{Variation of physical parameters along the jet}

The analysis performed on the IRS spectrum was also applied on the other
extracted knots in order to follow the variations of physical parameters
along the jet. In summary, we  performed the \feii\, line fit 
and used the \sii\, diagnostic analysis as described in the previous section for
deriving the A$_V$, $n_e$, and $T_e$ values. However, significant constraints on $x_e$ 
cannot  be derived along the jet since the limits on the 
involved line ratios are not stringent enough. 
Table 3 summarizes the obtained values, while Fig. 9 presents \feii\,
and \sii\, diagnostic diagrams compared with dereddened line ratios
that illustrate how the physical parameters vary from knot to knot. 

From Fig. 9 and Table 3 we see that electron density and temperature decrease
from the inner to the outer knots, a trend already observed in
jets from T Tauri stars (e.g., Lavalley-Fouquet et al. 2000, Podio et al. 2009, Maurri et al. 2014) and 
from Class I sources (Bacciotti \& Eisloffel 1999, Nisini et al. 2005, Podio et al. 2006). 

Garcia Lopez et al. (2008) derived the electron density along the HH34 jet
using the \feii\,1.64/1.60\um\, line ratio. 
They find a difference in density between the HVC and LVC,
i.e., the LVC density higher by a factor of two with respect to the HVC,
which we do not see in our data. This is probably due to the 
different extraction regions of the two data sets. In Garcia Lopez et al.,
their A6 knot comprises our IRS, blue~1, and blue~2 extractions (i.e., about 3\arcsec). Therefore, their
HVC density, averaged over a large region at progressively decreasing density,
is lower than our derived value by a factor of 5. In contrast, the LVC component is compact
and not resolved in both of the observations; therefore, the two density
values are more similar (2.2\,10$^4$ against 5\,10$^4$ \cmt). 

In general, we find density values higher by a factor
of $\sim$ 2 with respect to the Garcia Lopez et al. values. We think this 
difference is due to the different aperture
sizes and also to the use of a larger number of lines from different 
\feii\, levels sensitive to higher density conditions.

\begin{table*}
\caption[]{Derived physical parameters}
\vspace{0.5cm}
\begin{tabular}{rcccccccc}
\hline
       & $A_{V}$(mag) & \multicolumn{2}{c}{$n_e$(cm$^{-3}$)} &  \multicolumn{2}{c}{$T_e$(K)} &  \multicolumn{2}{c}{x$_e$}\\
 \cmidrule(lr){2-2}    \cmidrule(lr){3-4}  \cmidrule(lr){5-6}  \cmidrule(lr){7-9} 
       & \feii        & \feii   & \sii  &  \feii   & \sii  & \feii/\feiii & \oi/\oii & [\ion{N}{i}]/\nii \\
IRS-HV~LD$^a$&  9         & 5\,10$^4$    & 7\,10$^3$ - 2\,10$^4$& $>$15\,000  & $>$10\,000  & $<$0.35 & $<$0.3 & $<$0.1 \\
~~~~~~~~HD$^a$& $>$8      & $>$5\,10$^6$ & ...                  & $>$6000  & ...  & ... & ... & ...\\
IRS-LV &   9          & 5\,10$^4$ & 5\,10$^3$ - 2\,10$^4$   & 7000  & 6000-12\,000  &... & ... & ...\\
Blue 1 &   6.7        &  8\,10$^3$ & 3-4\,10$^3$            & 12\,000-15\,000  & 7000-10\,000  &... & ... & ...\\
Blue 2 &   5.8        & 4\,10$^3$ &2-3\,10$^3$              & 7000  & 6000-7000  &... & ... & ...\\
Blue 3 &   4.7        &2\,10$^3$ & $\sim$10$^3$             & 6000  & $<$15\,000 &... & ... & ...\\
\\[+5pt]  
\hline\\[-5pt]
\end{tabular}
\\
$^a$ High velocity - low density component\\
$^b$ High velocity - high density component\\
\end{table*}

\begin{figure}
\includegraphics[angle=0, width = 9cm,trim={0 6cm 0 6cm},clip ]{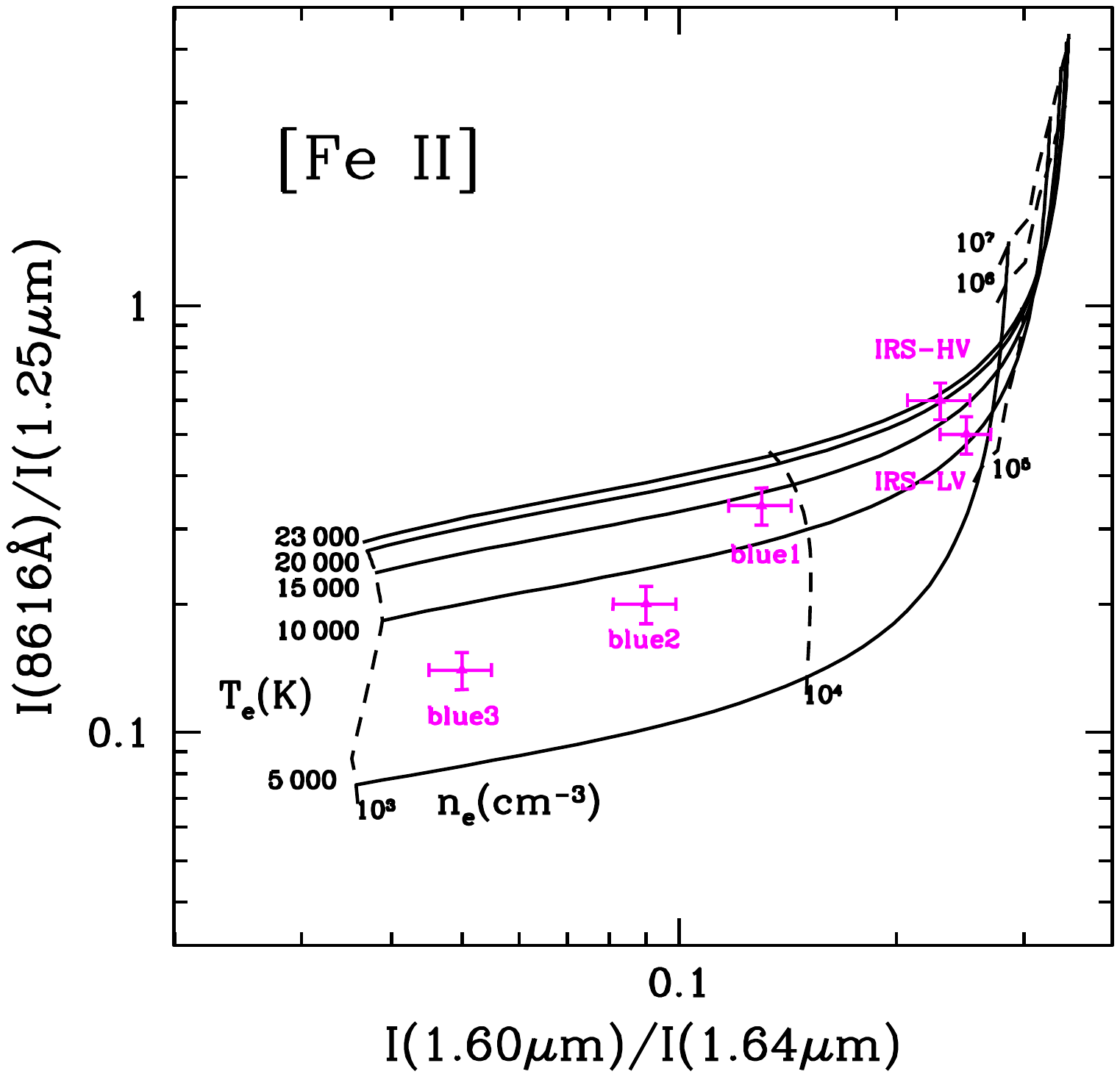}
\includegraphics[angle=0, width = 9cm, trim={0 6cm 0 6cm},clip]{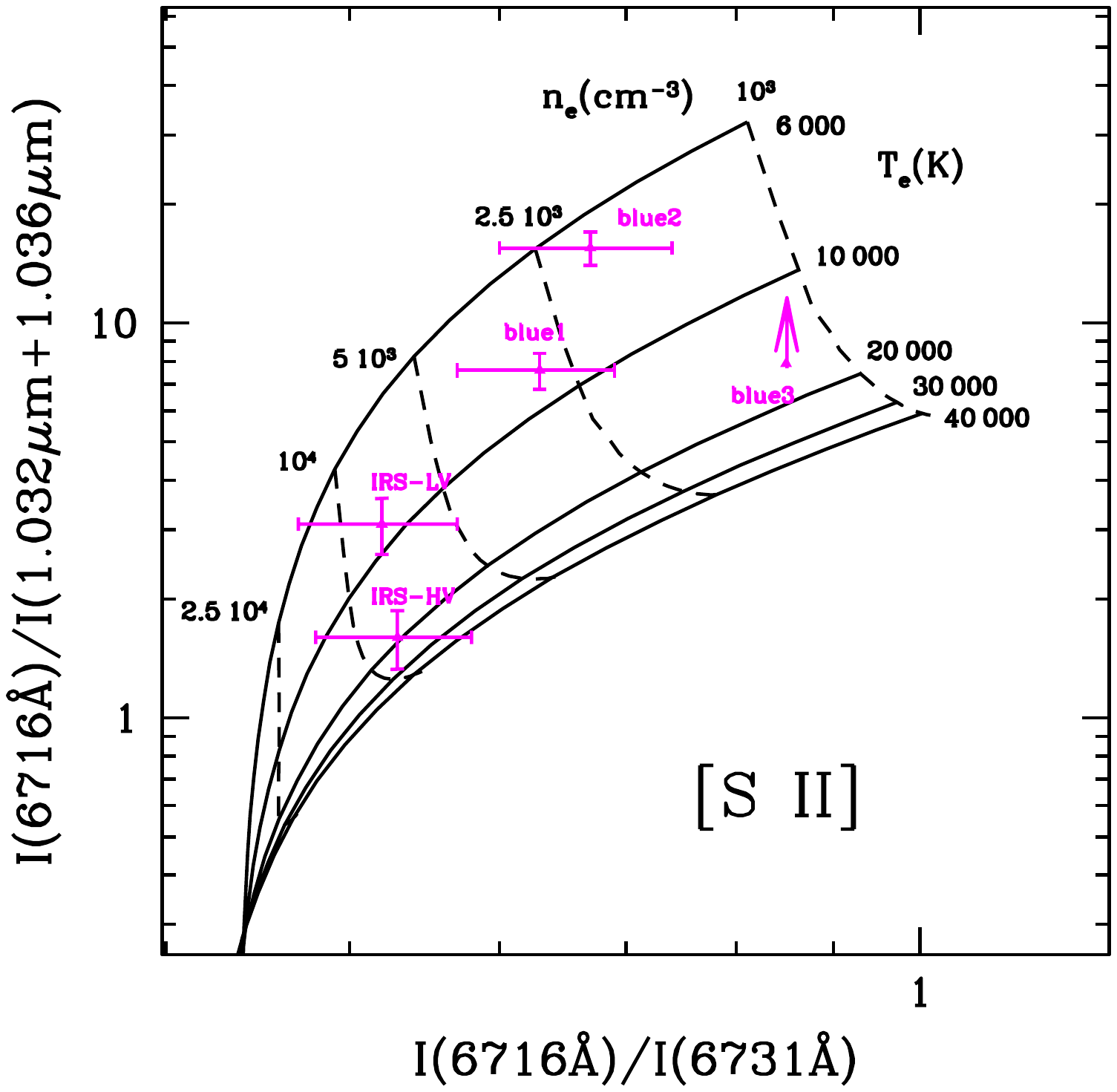}
      \caption{Diagnostic diagrams employing density and temperature sensitive 
      ratios of \feii\, and \sii\, transitions. Observed ratios, dereddened assuming
      the $A_{\rm V}$ values given in Table 3, are shown as purple symbols.
              }
         \label{fig.9}
   \end{figure}   
   
\subsection{Iron depletion}

As shown in section 4.1, intensity variations with velocities observed on the line profiles 
suggest that the abundances of refractory elements, like \feii, \crii, and \niii ,
decreases in the LVC as compared with the high velocity gas. 
These elements in gas phase can be underabundant with respect to solar values owing
to depletion on dust grains. In fact, a measure of the gas phase depletion of refractory
species such as Fe offers a direct indication of the presence of dust in the 
emitting region, and hence makes it possible to determine whether the outflowing material originates inside or outside the dust sublimation radius. 
Previous estimates of the Fe gas phase abundance in jets indicate depletion factors 
ranging between 20 to 80\% with respect to solar values (Nisini et al. 2002, 2005; Podio et al. 2006; Giannini et al. 2013) at distances larger than $\sim$ 400 AU from the source. 
Agra-Amboage et al. (2011) measured the iron depletion in the DG Tau jet 
down to 50 AU from the source. They found a depletion of
a factor $\sim 3$ in the high velocity component and a factor $\sim 10$ at velocities 
$<$ 100\kms . The HH34 observations probe the jet within $\sim$ 200 AU from the 
source and thus our iron abundance estimates can be compared reasonably well with
those of DG Tau. 

Iron depletion can be measured by comparing \feii\, fluxes with those of lines
from species that show no depletion onto grains. These reference lines
should be emitted in the same volume of gas as the \feii\, lines. 
To this end, infrared \feii\, lines have been in the past compared with IR [\ion{P}{ii}] 
lines in HH objects and jets (e.g., Nisini et al. 2005) as the IR transitions of the two species share
similar excitation conditions. In the optical, a suitable ratio is 
the \feii 8616\AA/\oi~6300~\AA\, since these two lines show 
similar PV diagrams and are close enough in wavelength to minimize errors induced
by A$_V$ uncertainties. The use of this ratio also implies   assuming that
oxygen is mostly neutral and iron mostly ionized, an assumption confirmed by the 
low ionization fraction derived.

In Fig. 10 we plot the \feii~8616~\AA/\oi~6300~\AA\, 
expected ratio as a function of the gas temperature, for the density value
derived in the HH34~IRS region. Different curves are plotted 
that assume a relative [Fe/O] solar abundance of 0.062 (Asplund et al. 2006)
and decreasing fractions of this value. The values derived from the 
observations in the HVC and LVC are shown with their errors.
The HVC is consistent with
the solar abundance value and $T_e \sim$15\,000 K.
For the LVC, a larger error is given on the \feii/\oi\, ratio that is due
to the contamination of the \oi\,6300\AA\, line by residuals 
of the telluric line subtraction. The gas phase [Fe/O] value derived for this component
implies Fe depletion of a factor between 2 and 8 with respect 
to solar.  

We therefore find results consistent with the DG Tau case where the \feii\,
depletion in the LVC is larger than in the HVC. 
A similar depletion pattern as a function of velocity has also been found
 from the analysis of other refractory species, e.g., Ca (e.g., Podio et al. 2009).
As discussed in Agra-Amboage
et al. (2011), this implies that the low velocity gas originates from a region
located at a larger distance from the star and/or from a region where
the dust has been less shock-reprocessed. At variance with the DG Tau case, however,
we do not see a significant iron depletion in the high velocity gas.

\begin{figure}
\includegraphics[angle=0, width = 9cm,trim={0 6cm 0 6cm},clip ]{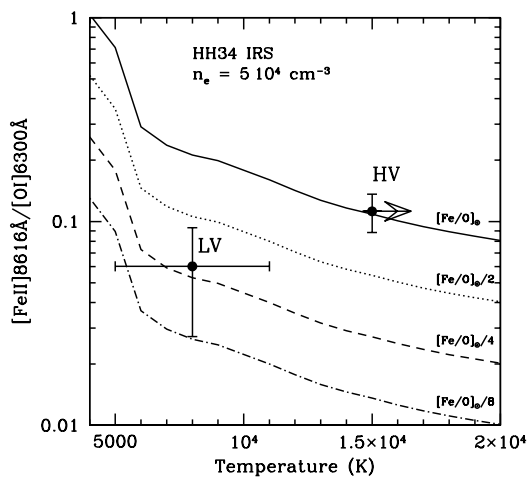}
      \caption{\feii~8616~\AA/\oi~6300~\AA\, ratio plotted against the gas temperature
      for a density of 5\,10$^4$ \cmt . The solid curve shows predictions 
      assuming an [Fe/O] solar gas phase abundance of 0.062 (Asplund et al. 2006).
      Other curves are plotted for decreasing values of the [Fe/O] abundance. 
      Data points correspond to the values measured in the HVC and LVC 
      toward HH34~IRS, and indicate an [Fe/O] abundance ratio close to solar
      for the HVC and a factor between 2 and 8 below solar for the LVC. 
              }
         \label{fig.10}
   \end{figure}

\section{Mass accretion rate}

The on-source spectrum of HH34 presents several optical and IR permitted lines 
that are often used as proxies for the accretion luminosity ($L_{acc}$)
in young stars. Among the
brightest are the HI (H$\alpha$, Pa$\beta$,  Br$\gamma$), 
OI (8446.4\AA), and Ca II (8498\AA, 8542\AA, 8662\AA) transitions. 
Although the real origin of these lines -- whether in accretion columns and spots,
or in ionized winds -- is still uncertain, it has been shown that their intrinsic 
luminosity scales with $L_{acc}$ in T Tauri stars (Herczeg 
\& Hillenbrand 2008; Alcal\'a et al. 2014). As such they have been widely used
to indirectly derive $L_{acc}$ and consequently the mass accretion rate 
(\.M$_{acc}$) in sources with known stellar parameters (e.g., Antoniucci et al. 2014b).
Using the assumption that the same relationships between $L_{line}$ vs $L_{acc}$
also hold  in the more embedded class I sources, we can estimate \.M$_{acc}$
 in HH34~IRS from the dereddened line luminosities.
The main uncertainty in this procedure is the knowledge of the extinction value
in regions close to the source where the permitted lines originate.

A direct estimate of the on-source extinction value can be gathered by the 
depth of absorption bands detected in the Spitzer-IRS on-source spectrum (see Fig. 2).
 In particular, the  silicate absorption
feature at 9.7\um\ can be used by adopting the 
relationship A$_V$/$\tau_{9.7}$ = 18.5$\pm$1.5 mag from Mathis (1998). 
This gives an  A$_V$ value between 30 and 33 mag. 
A high value of A$_V$ is also estimated by Antoniucci et al. (2008),
who applied a self-consistent method to simultaneously
derive A$_V$, $L_{acc}$, and $L_{*}$ in HH34~IRS from the source bolometric
luminosity, IR magnitudes, and Br$\gamma$ flux. They infer an A$_V$ value close to
50 mag and a \.M$_{acc}$ of 4.1\,10$^{-6}$ M$_{\odot}$\,yr$^{-1}$.
Hence the permitted lines might originate in a region more deeply embedded 
than the jet region where the forbidden lines are located.  

Taking advantage of the wide wavelength range  covered by our data,  in principle  we can
derive simultaneously A$_V$ and \.M$_{acc}$ 
by considering the A$_V$ that gives consistent \.M$_{acc}$ values from the different dereddened
line luminosities (in particular H$\alpha$, \ion{O}{I}8446\AA, \ion{Ca}{II}8662\AA, Pa$\beta$, and Br$\gamma$). Applying this method, and using  the empirical relationships between $L_{line}$ and $L_{acc}$ 
derived by Alcal\'a et al. (2014), we find A$_V$ = 7 mag and \.M$_{acc}$ = (8$\pm$7)\,10$^{-9}$ M$_{\odot}$\,yr$^{-1}$ (assuming M$_{*}$ = 0.5 M$_{\odot}$ and considering absolute calibration errors on the line luminosity). 

The  A$_V$ derived in this way is much lower than that measured from the silicate band and
the \.M$_{acc}$ value is two order of magnitudes lower than the jet mass flux rate.
In addition, following the arguments of Antoniucci et al. (2008), such a low  mass accretion 
rate would imply that the bolometric luminosity of the central source
(12-20 L$_\odot$) is not accretion dominated, but rather
dominated by the stellar photosphere. However, this is not consistent with the 
source IR photometry and A$_V$ = 7 mag value for any 
central star spectral type.

A possible explanation for this discrepancy could be that a significant fraction of the 
permitted line emission originating very close to the star is  seen here as emission scattered in the 
cavity excavated envelope. The net effect of scattering would be to enhance 
the line emission at shorter wavelengths with respect to the IR lines, with a 
wavelength dependent law different from the adopted reddening law.
We therefore conclude  that the relationships between accretion luminosity and permitted
optical line luminosity derived for T Tauri stars cannot be directly extended to class I sources 
embedded in reflection nebulae.

An alternative way to indirectly estimate the mass accretion rate is to use empirical
correlations between \.M$_{acc}$ and the luminosity of forbidden lines.
For example, Herczeg \& Hillenbrand (2008) found 
a correlation between the \oi\, 6300\AA\, luminosity and the accretion luminosity
in T Tauri stars. This relationship has been used to infer the 
mass accretion rates in embedded objects or in edge-on-disk T Tauri where
the permitted line emission is partially hidden at a direct view (Whelan et al. 2014,
Riaz et al. 2015).
However, the \oi\, emission in the Herczeg \& Hillenbrand sample was dominated 
by the low velocity wind component. 
Nisini et al. (in prep.) revised this correlation for a sample of  
CTTs of the Lupus and Cha clouds, considering only the 
\oi\, 6300\AA\, luminosity in the HVC. They found the following
relationship:  log ($L_{[\ion{O}{i}]\rm{HVC}}$) = 0.8\,log (\.M$_{acc}$) + 1.8. 
Assuming that the same relationship holds also for class I sources,  
for HH34 IRS we derive \.M$_{acc}$ = (7.5$\pm$ 4)\,10$^{-6}$ M$_{\odot}$\,yr$^{-1}$. 
This is close to the value originally derived by Antoniucci et al. (2008) and
implies a \.M$_{jet}$/\.M$_{acc} \sim 0.1 $, i.e., well within the range 
expected for magneto-centrifugally jet launching models (Ferreira et al. 2006).

\section{Discussion and conclusions}

In this section we summarize and discuss the main results found from the
analysis of the X-shooter spectrum of the HH34~IRS micro-jet.

\subsection{ Kinematics vs excitation}

Thanks to the large spectral coverage and sensitivity reached by the X-shooter 
observations, we were able to detect lines spanning
a wide range of excitation energies (from $\sim$ 8000 to $\sim$ 31\,000 cm$^{-1}$)
and critical densities (from $\sim$ 10$^3$ to $\sim$ 10$^7$ cm$^{-3}$).
We find  that  PV diagrams of lines with different excitation conditions are remarkably
similar. The peak velocity and the FWHM of the  HVC  
is the same for all lines, even though -- at least in the on-source position -- 
high excitation lines seem to trace a denser gas. Hence density stratifications
are not associated with different velocity components. In particular, we
do not see evidence at our spectral resolution that high velocity material
located close to the jet axis, has higher density than the jet
streamlines at lower velocities on the border. 
This is in contrast with the onion-like structure predicted by the theoretical model and 
confirmed by high resolution observations of CTT jets (e.g., Maurri et al. 2014), 
and may be due to the lack of angular resolution: in regions more distant from the driving source 
(like those probed here), shocks may have already destroyed the jet's onion-like structure.
With regard to the spatial variations, all  transitions peak at the same offsets from the source
with the exception of the \sii\, and \caii\, lines, which -- having critical densities $<$ 5\,10$^3$ cm$^{-3}$
-- are probably quenched in the high density knots close to the star. 
We therefore conclude that shock layers at different excitation behind the shock front 
are not spatially resolved.

\subsection {Physical parameters, mass loss rate, and comparison with properties of CTT stars}

Our study presents the first estimates of the full set of 
physical parameters ($n_e$, $T_e$, x$_e$, $n_{H}$) in the initial 200 AU of 
a jet from an embedded class I source.
The values found in this inner jet are $T_e \ga $ 15\,000 K, $n_e \sim$ 5\,10$^4$ \cmt,
and fractional ionization x$_e <0.1$, which implies a total density $n_H >$ 5\,10$^5$ \cmt.
A gas component whose density is  at least two order of magnitudes higher  is also
suggested by the intensity of the \feii\, lines at higher excitation. However, this component does not
contribute to more than 1\% of the jet mass if we assume for it the same
fractional ionization as the lower density gas.

Although the results derived from this single source cannot be generalized
to all the class I objects, it is  instructing to compare the derived parameters with 
values typically found in CTTs jets, especially for  the fractional ionization
and total density,  the parameters derived here for the first time.

The total density estimated in the HH34 jet ($n_H >$ 5\,10$^5$ \cmt)
 is at the upper boundary of the typical values
found in CTTs on the same spatial scales of our observations. Values in the range 
10$^4$-10$^5$\cmt\ are derived in DG Tau, HH30, and RW Aur
(Maurri et al. 2014; Hartigan \& Morse et al. 2007; Melnikov et 
al. 2009), which are among the most active CTTs.

Fractional ionizations in CTT jets have strong gradients in the initial $\sim$ 150 AU
from the source. In DG Tau, for example, the ionization fraction
increases from $\sim$0.07 at the jet base to up to 0.6-0.8 above 150 AU (Maurri
et al. 2007). Similarly, in HH30, x$_e$ is $\la$0.1 within 50 AU of the source,
and then progressively increases up to $\sim$0.3 at about 200 AU (Hartigan \& Morse et al. 2007). 
We don't have the spatial resolution to see x$_e$ variations in this
inner jet sections and we measure x$_e <$ 0.1 as an average value of the region
where such strong gradients are expected. Podio et al. (2006) have also shown that 
x$_e$ in HH34 never rises above 0.1 at large distances. 
Such low values of fractional ionizations, however,  are  found in the jet from RW Aur,
where x$_e \sim$  0.1 has been measured along  the entire micro-jet up to distances of few arcsec
(Melnikov et al. 2009).
In gas excited by shocks, low values of fractional ionization are expected
either for low shock velocity or for high pre-shock density. Since the HH34 jet velocity
and velocity spread are similar to other T Tauri stars where higher ionization has been
found, the low x$_e$ value is probably a direct consequence of the high density in the jet. 

The mass flux rate derived from the estimated value of the total density is 
 \.M$_{jet} >$ 5\,10$^{-7}$ M$_{\odot}$\,yr$^{-1}$.
This is at least an order of magnitude higher than values
measured on active T Tauri stars (see, e.g., Ellerbroek et al. 2013).
We therefore conclude that the class I HH34~IRS jet shares many kinematical  and physical 
   properties with the most active CTTs stars, but its density and mass flux is at least 
   an order of magnitude higher. 

\subsection{Origin of the LVC}
 
The LVC at the source position is detected in all the brightest lines and it is
 blue-shifted by $\sim$ 30-50~\kms . From the diagnostic analysis it results that the LVC
 shares the same electron density and extinction as the base of the HVC, but has a lower
 temperature,  on the order of $\sim$ 7000 K. 
 Forbidden line emission components centered close to systemic velocity are features commonly observed in CTT stars (e.g., Hartigan et al. 1995; Natta et al. 2014).
 The typical peak velocities of CTT LVCs are in the range $\sim$  5-20\kms\ (thus closer to systemic
 velocity than the HH34 LVC) and they are usually brighter than the corresponding HVC emission, 
 even in sources known to drive prominent jets (e.g., Hartigan et al. 1995). 
 As we estimate a similar extinction between the HVC and LVC, the weakness of the LVC cannot be attributed
 to an origin in a region more embedded than the high velocity jet gas emission.
 The origin of the LVC is unclear: in CTTs it has been attributed either to low velocity magnetically driven disk-winds or to winds  photo-evaporated by UV or X-ray photons from the central star. 
 An origin in photo-evaporated wind is very unlikely in the HH34 IRS case, 
 because the peak velocity appears much more blue-shifted than in the line profiles
 predicted by models (e.g., Ercolano \& Owen 2010) and because it is conceivable 
 that any stellar high energetic photon would be absorbed by the dense jet before
 being able to photoevaporate disk material. It is therefore more likely that the LVC originates 
 from a slow magnetically accelerated disk-wind. Different flow acceleration models 
 do indeed predict the formation of compact slow winds. 
 
 Romanova et al. (2009), for example, provide numerical simulations of outflows launched from the interaction region between the disk and a rotating magnetized star, which in principle could explain the presence of a LVC. This model  predicts the formation of  a fast axial-symmetric jet and of a conical inner wind powered by the magnetic
 pressure force. Romanova et al. (2009) applied their simulations to different types of stars.
 For the case of class I sources relevant here, the protostar is assumed to be rapidly rotating
 causing the formation of an energetic jet that carries away a relevant amount of angular
 momentum and of a slow conical wind with poloidal velocities up to 50 \kms . Thus, 
 the model could explain the kinematics we observe in our data. However the conical wind
 is predicted to be at higher density with respect to the jet, which is contradicted by our
 analysis. In addition, the wind is formed in the inner disk region close to the star: this is
 not consistent with the \feii\, depletion we observe in the LVC, which suggests a formation
 in a disk region where the dust has not sublimated yet.
 
 A different class of models are those that consider an outflow originating from a
 magnetized disk (e.g., Konigl \& Pudritz 2000; Ferreira et al. 2006). In these models the 
 wind is ejected from an extended range of radii and thus they naturally  predict 
 components at different velocities as arising in flows from different streamlines at
 decreasing poloidal velocity across the disk. Extended MHD disks can also take into account 
 the \feii\, depletion pattern as the gas at low velocity is likely ejected from large disk
 radii, outside the dust sublimation radius. This class of models also predicts that 
  the particle density should decrease along the radial streamlines, a feature in apparent
  contradiction with our finding of constant electron density. 
  However, the total density could still be consistent with the predicted
  pattern if the ionization fraction in the LVC is smaller than in HVC. In this respect, we point out that 
a decrease in the fractional ionization with velocity has been observed in T Tauri stars 
(e.g., Maurri et al. 2014). 

\subsection{Mass accretion rate in class I sources}

The luminosity of optical/IR permitted lines is routinely used to measure the accretion 
luminosity and thus accretion rates in CTT stars adopting empirical relationships
calibrated on samples of stars with known accretion luminosity, derived, for example,
from the UV veiling and Balmer jump. In embedded sources, where the UV 
spectral region is not directly accessible, the possibility to use these empirical relationships
would provide a very important tool to obtain information on the accretion properties 
of the sources. However, the analysis we have performed on source HH34~IRS
 shows that the optical lines are likely seen through scattered light and thus
their intrinsic luminosity cannot be retrieved by adopting a standard reddening law. 
Methods that self-consistently infer both \.M$_{acc}$ and
 A$_V$ combining Br$\gamma$ line luminosity and stellar parameters,  as
 adopted in Antoniucci et al. (2008), seem better able to infer the accretion properties 
 for these  objects. This method, however, relies on a good knowledge of
 the source bolometric luminosity.
 
Empirical relationships correlating \.M$_{acc}$ with the luminosity of forbidden
lines, such as the \oi\, line, can also be used to estimate the mass accretion
rate. These relationships, calibrated on CTT stars, hinges on the connection between
\oi\, luminosity and mass ejection rate, which in turn indirectly depends on the mass accretion
rate. Therefore, its goodness eventually relies on the assumption that 
the HH34~IRS jet shares a similar mass ejection efficiency to that of  T Tauri stars.

Ultimately, however, for better accretion proxy in class I sources not contaminated by scattered light and less dependent on extinction, it is necessary to go into the mid-IR regime. Examples of 
studies toward that direction are those of Salyk et al. (2013) who used the Pf$\beta$ 
luminosity, and Rigliaco et al. (2015), who explored the potential of the HI(7-6) at
12.37$\mu$m.



      

%


\begin{acknowledgements}
This project was financially supported by the PRIN INAF 2012 ``Disks, jets and the dawn of planets''.
L.P. has received funding from the European Union Seventh Framework Programme (FP7/2007$-$2013) under grant agreement No. 267251.
\end{acknowledgements}


\begin{appendix}

\section{Tables of line fluxes.}

\onecolumn
\begin{longtable}{clccccccc}
\caption[]{Lines observed in the on-source extraction: the HV and LV components}\\
\hline
$\lambda _{air}$ & Ion & Upper & Lower &$E_{up}$ &  Flux (HV) & Flux (LV) & $\Delta$(Flux) & \\
(\AA) &  & level & level & (cm$^{-1}$) &  \multicolumn{3}{c}{(10$^{-17}$\,erg\,s$^{-1}$\,cm$^{-2}$)} & \\[+5pt]  
\endfirsthead
\caption{Continued.} \\
\hline
$\lambda _{air}$ & Ion & Upper & Lower &$E_{up}$ &  Flux (HV) & Flux (LV) & $\Delta$(Flux) & \\
(\AA) &  & level & level & (cm$^{-1}$) &  \multicolumn{3}{c}{(10$^{-17}$\,erg\,s$^{-1}$\,cm$^{-2}$)} &\\[+5pt]  
\endhead
\hline\\[-5pt] 
\multicolumn{9}{c}{Oxygen} \\
\hline\\[-5pt] 
6300.3 & \oi   & $^1\!D_{2}$   & $^3\!P_{2}$   & 15867.8 &  37.5  &  12.2  &    0.1& \\
6363.7 & \oi   & $^1\!D_{2}$   & $^3\!P_{1}$   & 15867.8 &  13.8  &   5.3  &    0.1& \\
7319.9 & \oii  & $^2\!P_{3/2}$ & $^2\!D_{5/2}$ &40468.0  &   $<$0.5  &    ...  &   ...& \\
7330.7 & \oii  & $^2\!P_{3/2}$ & $^2\!D_{3/2}$ &40468.0  &   $<$0.5  &    ...  &   ...& \\
\hline\\[-5pt] 
\multicolumn{9}{c}{Sulphur} \\
\hline\\[-5pt] 
6716.4 & \sii & $^2\!D_{5/2}$ & $^4\!S_{3/2}$ & 14884.7   &   8.9  &   7.5  &    0.1& \\
6730.8 & \sii & $^2\!D_{3/2}$ & $^4\!S_{3/2}$ & 14852.9   &  17.4  &  14.6  &    0.1& \\
10286.7 &  \sii & $^2\!P_{3/2}$ & $^2\!D_{3/2}$ & 24571.5 &  47.6  &  14.8  &    0.7& \\
10320.5 &  \sii & $^2\!P_{3/2}$ & $^2\!D_{5/2}$ & 24571.5 &  61.5  &  25.6  &    0.7& \\
10336.4 &  \sii & $^2\!P_{1/2}$ & $^2\!D_{3/2}$ & 24524.8 &  38.4  &  16.8  &    0.1& \\
10370.5 &  \sii & $^2\!P_{1/2}$ & $^2\!D_{5/2}$ & 24524.8 &   9.8  &  30.0  &    0.1& \\
\hline\\[-5pt] 
\multicolumn{9}{c}{Nitrogen} \\
\hline\\[-5pt] 
5197.9 & [\ion{N}{i}] & $^2\!D_{3/2}$ & $^4\!S_{3/2}$ & 19233.1    & 1.0  &    ...  &    0.2& \\
6548.0 & \nii & $^1\!D_{2}$ & $^3\!P_{1}$ & 15316.1                &  0.6  &    ...  &    0.1& \\
6583.4 & \nii & $^1\!D_{2}$ & $^3\!P_{2}$ & 15316.1                & 3.6  &    ...  &    0.1& \\
10397.7 & [\ion{N}{i}]  & $^2\!P_{3/2}$ & $^2\!D_{5/2}$ & 28839.3  &66.8  &  42.0  &    0.7& \\
10398.1     &      & $^2\!P_{1/2}$ & $^2\!D_{5/2}$ & 28838.9       &       &       &     & \\
10407.1 & [\ion{N}{i}]  & $^2\!P_{3/2}$ & $^2\!D_{3/2}$ & 28839.3  &49.2  &  22.6  &    0.7& \\
10407.5     &      & $^2\!P_{1/2}$ & $^2\!D_{3/2}$ &  28838.9      &       &       &     & \\
\hline\\[-5pt] 
\multicolumn{9}{c}{Iron} \\
\hline\\[-5pt] 
5158.0 & \feii & $b^4\!P_{3/2}$ & $a^4\!F_{7/2}$ & 21812.0   &   3.8  &    ...  &   0.2& \\
5261.6 & \feii & $a^4\!H_{11/2}$ & $a^4\!F_{7/2}$ &  21430.3 &   2.2  &    ...  &   0.2& \\
5273.3 & \feii & $b^4\!P_{5/2}$ & $a^4\!F_{9/2}$ & 20830.5   &   0.6  &    ...  &   0.2& \\
5333.6 & \feii & $a^4\!H_{9/2}$ & $a^4\!F_{5/2}$ & 21581.6   &   2.5  &    ...  &   0.2& \\
5376.4 & \feii & $a^4\!H_{7/2}$ & $a^4\!F_{3/2}$ & 21711.9   &   0.7  &    ...  &   0.2& \\
5527.3 & \feii & $a^2\!D_{5/2}$ & $a^4\!F_{7/2}$ & 20516.9   &   1.1  &    ...  &   0.2& \\
7155.1 & \feii & $a^2\!G_{9/2}$ & $a^4\!F_{9/2}$ & 15844.6   &  27.8  &   4.6  &    0.2& \\
7172.0 & \feii & $a^2\!G_{7/2}$ & $a^4\!F_{7/2}$ & 16369.4   &   7.5  &   1.3  &    0.1& \\
7388.1 & \feii & $a^2\!G_{7/2}$ & $a^4\!F_{5/2}$ & 16369.4   &   4.3  &   6.2  &    0.1& \\
7452.5 & \feii & $a^2\!G_{9/2}$ & $a^4\!F_{7/2}$ & 15844.65  &  12.4  &   2.5  &    0.1& \\
7686.9 & \feii & $a^4\!P_{3/2}$ & $a^6\!D_{5/2}$ & 13673.2   &   3.9  &   1.2  &    0.2& \\ 
8616.9 & \feii & $a^4\!P_{5/2}$ & $a^4\!F_{9/2}$ & 13474.4   &  69.5  &  20.9  &    0.1& \\
8891.9 & \feii & $a^4\!P_{3/2}$ & $a^4\!F_{7/2}$ & 13673.2   &  29.6  &   8.3  &    0.1& \\
9033.4 & \feii & $a^4\!P_{1/2}$ & $a^4\!F_{5/2}$ & 13904.8   &   9.6  &   4.2  &    0.4& \\
9051.9 & \feii & $a^4\!P_{5/2}$ & $a^4\!F_{7/2}$ & 13474.4   &  20.7  &   6.2  &    0.4& \\
9226.6 & \feii & $a^4\!P_{3/2}$ & $a^4\!F_{5/2}$ & 13673.2   &  13.9  &  18.9  &    0.2& \\
9267.5 & \feii & $a^4\!P_{1/2}$ & $a^4\!F_{3/2}$ & 13904.8   &  17.3  &   6.9  &    0.2& \\
9399.0 & \feii & $a^4\!P_{5/2}$ & $a^4\!F_{5/2}$ & 13474.4   &   7.3  &    ...  &   1.0& \\
12485.4 & \feii & $a^4\!D_{5/2}$ & $a^6\!D_{7/2}$ & 8391.9   &  25.0  &  10.2  &    0.5& \\
12521.3 & \feii & $a^4\!D_{1/2}$ & $a^6\!D_{3/2}$ & 8846.8   &  19.0  &   7.4  &    0.5& \\
12566.8 & \feii & $a^4\!D_{7/2}$ & $a^6\!D_{9/2}$ & 7955.3   & 580.7  & 267.2  &    0.5& \\
12703.4 & \feii & $a^4\!D_{1/2}$ & $a^6\!D_{1/2}$ & 8846.8   &  77.9  &  45.5  &    0.7& \\
12787.7 & \feii & $a^4\!D_{3/2}$ & $a^6\!D_{3/2}$ & 8680.45  & 118.1  &  65.1  &    0.7& \\
12942.7 & \feii & $a^4\!D_{5/2}$ & $a^6\!D_{5/2}$ & 8391.94  & 169.7  &  75.8  &    0.4& \\
12977.7 & \feii & $a^4\!D_{3/2}$ & $a^6\!D_{1/2}$ & 8680.45  &  57.0  &  19.3  &    0.4& \\
13205.5 & \feii & $a^4\!D_{7/2}$ & $a^6\!D_{7/2}$ & 7955.30  & 224.4  &  86.5  &    0.4& \\
13277.1 & \feii & $a^4\!D_{5/2}$ & $a^6\!D_{3/2}$ & 8391.94  &  85.3  &  66.3  &    0.4& \\
15334.7 & \feii & $a^4\!D_{5/2}$ & $a^4\!F_{9/2}$ & 8391.94  & 308.5  & 148.5  &    0.9& \\
15994.7 & \feii & $a^4\!D_{3/2}$ & $a^4\!F_{7/2}$ & 8680.45  & 272.2  & 143.7  &    0.7& \\
16435.4 & \feii & $a^4\!D_{7/2}$ & $a^4\!F_{9/2}$ & 7955.3   &1296.5  & 519.7  &    0.7& \\
16637.6 & \feii & $a^4\!D_{1/2}$ & $a^4\!F_{5/2}$ & 8846.8   & 159.9  &  80.0  &    0.7&  \\
16768.7 & \feii & $a^4\!D_{5/2}$ & $a^4\!F_{7/2}$ & 8391.9   & 284.0  & 159.9  &    0.7& \\
17111.3 & \feii & $a^4\!D_{3/2}$ & $a^4\!F_{5/2}$ & 8680.4   & 104.5  &  52.7  &   1.0& \\
17484.2 & \feii & $a^4\!P_{3/2}$ & $a^4\!D_{7/2}$ & 7955.30  &  17.9  &    ...  &    0.7& \\
17971.1 & \feii & $a^4\!D_{3/2}$ & $a^4\!F_{3/2}$ & 8680.4   & 211.1  & 102.8  &    0.7& \\
18000.2 & \feii & $a^4\!D_{5/2}$ & $a^4\!F_{5/2}$ & 8391.9   & 250.2  & 167.1  &   1.8& \\
18093.9 & \feii & $a^4\!D_{7/2}$ & $a^4\!F_{5/2}$ & 7955.3   & 359.0  & 134.6  &   1.8& \\
18954.0 & \feii & $a^4\!D_{5/2}$ & $a^4\!F_{3/2}$ & 8391.9   &  55.9  &  26.3  &  12.5& \\
20151.2 & \feii & $a^2\!H_{9/2}$ & $a^2\!G_{9/2}$ & 20805.8  & 143.2  &    ... &   5.0& \\
20460.1 & \feii & $a^2\!P_{3/2}$ & $a^4\!P_{5/2}$ & 18360.6  & 103.7  &  19.4  &   5.0& \\
21327.7 & \feii & $a^2\!P_{3/2}$ & $a^4\!P_{3/2}$ & 18360.6  &  66.0  &  18.8  &   3.2& \\
22237.6 & \feii & $a^2\!H_{11/2}$ & $a^2\!G_{9/2}$ & 20340.3 & 112.7  &  44.7  &   3.5& \\
5270.4 & \feiii & $^3\!P_{2}$ & $^5\!D_{3}$ & 19404.8        &   $<$6.0  &    ...  &   ...&   \\
\hline\\[-5pt] 
\multicolumn{9}{c}{Calcium} \\
\hline\\[-5pt] 
7291.4 & \caii & $^2\!D_{5/2}$ & $^2\!S_{1/2}$ & 13710.8            &  10.1  &    ...  &  0.1& \\
7323.8 & \caii & $^2\!D_{3/2}$ & $^2\!S_{1/2}$ & 13650.1            &   7.0  &    ... &   0.1& \\
8542.0$^a$ & \ion{Ca}{ii} & $^2\!P_{3/2}$ & $^2\!D_{5/2}$ & 25414.4     &6.1  &   ... & 0.2 &\\
8662.1$^a$ & \ion{Ca}{ii} & $^2\!P_{1/2}$ & $^2\!D_{3/2}$ & 25191.5 &   2.9  &    ... &   0.1& \\
\hline\\[-5pt] 
\multicolumn{9}{c}{Other forbidden lines} \\
\hline\\[-5pt] 
7377.8 & \niii & $^2\!F_{7/2}$ & $^2\!D_{5/2}$ & 13550.3     &  17.7  &    ...  &    0.1 & \\
7411.6 & \niii & $^2\!F_{5/2}$ & $^2\!D_{3/2}$ & 14995.5     &   3.9  &    ...  &    0.1& \\
8301.0 & \niii & $^2\!F_{7/2}$ & $^2\!D_{3/2}$ & 13550.3     &   2.3  &    ...  &    0.2& \\
19387.7 & \niii & $^2\!F_{7/2}$ & $^4\!F_{9/2}$ & 13550.3    & 305.0  &    ...  &  12.5& \\
8000.0 & \crii & $a^6\!D_{9/2}$ & $a^6\!S_{5/2}$ & 12496.4   &   6.1  &    ...  &    0.2& \\
8125.3 & \crii & $a^6\!D_{7/2}$ & $a^6\!S_{5/2}$ & 12303.8   &   5.4  &    ...  &    0.1& \\
8308.4 & \crii & $a^6\!D_{3/2}$ & $a^6\!S_{5/2}$ & 12032.5   &   3.1  &    ...  &    0.2& \\
8357.6 & \crii & $a^6\!D_{1/2}$ & $a^6\!S_{5/2}$ & 11961.8   &   2.6  &    ...  &    0.2& \\
8729.9 & \crii & $a^4\!F_{3/2}$ & $a^4\!D_{3/2}$ & 31082.9   &   9.7  &    ...  &    0.1& \\
9824.1 & [\ion{C}{i}] &  $^1\!D_{2}$ & $^3\!P_{1}$ & 10192.6 &  64.0  &    ...  &    0.6& \\
9850.2 & [\ion{C}{i}] &  $^1\!D_{2}$ & $^3\!P_{2}$ & 10192.6 & 115.9  &    ...  &    0.6& \\
\hline\\[-5pt] 
\end{longtable}
\begin{longtable}{clcccccc}
\caption[]{Lines observed in the on-source extraction: permitted lines}\\
\hline
$\lambda _{air}$ & Ion & Upper & Lower &$E_{up}$ &  Flux & $\Delta$(Flux) & \\
(\AA) &  & level & level & (cm$^{-1}$) &  \multicolumn{2}{c}{(10$^{-17}$\,erg\,s$^{-1}$\,cm$^{-2}$)} &\\[+5pt]  
\endfirsthead
\caption{Continued.} \\
\hline
$\lambda _{air}$ & Ion & Upper & Lower &$E_{up}$ &  Flux & $\Delta$(Flux)&  \\
(\AA) &  & level & level & (cm$^{-1}$) &  \multicolumn{2}{c}{(10$^{-17}$\,erg\,s$^{-1}$\,cm$^{-2}$)} &\\[+5pt]  
\endhead
\hline\\[-5pt] 
6562.8 & \ha & 3 & 2 & 97492.3 & 25.3 & 0.1 &\\
8446.4 & \ion{O}{i} &  $^3\!P$ & $^3\!S$ & 88631.3                  &2.1  &    0.2&\\
8498.0 & \ion{Ca}{ii} & $^2\!P_{3/2}$ & $^2\!D_{3/2}$ & 25414.4     &96.2  &   0.2 &\\
8518.0 & \ion{He}{i}  & $^1\!P_{1}$ & $^1\!S_{0}$ & 196601          &12.6  &   0.2&\\
8542.0$^a$ & \ion{Ca}{ii} & $^2\!P_{3/2}$ & $^2\!D_{5/2}$ & 25414.4     &97.8  &   0.2 &\\
8662.1$^a$ & \ion{Ca}{ii} & $^2\!P_{1/2}$ & $^2\!D_{3/2}$ & 25191.5 &86.5  &   0.1 &\\
10938.0 &  Pa$\delta$ & 6 & 3 & 106632.1 & 31.6  &   0.4&\\
12818.0 & Pa$\beta$ & 5 & 3 & 105291.6 & 338.2  &    0.7&\\
15880.5 & Br 14 & 4 & 14 & 109119.1      &  41.3  &   0.9 &\\
16109.3 & Br 13 & 4 & 13 & 109029.7      &  45.1  &   0.7 &\\
16407.2 & Br 12 & 4 & 12 & 108917.1      &  58.4  &   0.7 &\\
16806.5 & Br 11 & 4 & 11 & 108772.3      &  79.3  &   0.7 &\\
17362.1 & Br 10 & 4 & 10 & 108581.9      &  75.1  &   1.0 &\\
18174.1 & Br 9  & 4 & 9 & 108324.7       & 176.7  &   3.5&\\
18751.0 & Pa $\alpha$ & 3 & 4 & 102823.9 &2104.5  &   3.5&\\
20601.7 &  \ion{He}{i}  & $^3\!D_{1}$ & $^3\!P_{2}$ & 196069.7 & 43.5  &   5.0&\\
21655.3 & Br $\gamma$ & 4 & 7 & 107440.4 & 380.7  &   1.2 &\\
22056.4 & \ion{Na}{i} &  $^2\!P_{3/2}$ & $^2\!S_{1/2}$ & 30272.5 & 43.2  &   3.5&\\
22083.7 & \ion{Na}{i} &  $^2\!P_{1/2}$ & $^2\!S_{1/2}$ & 30266.9 & 46.0  &   3.5 &\\
\end{longtable}
\begin{longtable}{clcccc}
\caption[]{Lines observed on-source: molecular lines}\\
\hline
$\lambda _{air}$ & Molecule & term & Flux & $\Delta$(Flux) & \\
(\AA) &  & &\multicolumn{2}{c}{(10$^{-17}$\,erg\,s$^{-1}$\,cm$^{-2}$)} &\\[+5pt]  
\endfirsthead
\hline\\[-5pt] 
17480.34 & \htwo & 1-0 S(7) &  16.9  &   1.0 &\\
19575.6 &  \htwo & 1-0 S(3) & 177.5  &   5.0&\\
20337.5 &  \htwo &  1-0 S(2)&  93.0  &   3.7&\\
21218.3 &  \htwo & 1-0 S(1) &1164.5  &   3.2&\\
22233.0 & \htwo & 1-0 S(0)  &  85.2  &   3.5&\\
24065.9 &  \htwo & 1-0 Q(1) & 460.0  &  12.5&\\
24134.3 &  \htwo & 1-0 Q(2) & 174.2  &  12.5&\\
24237.3 & \htwo &  1-0 Q(3) & 328.0  &  12.5&\\
24374.9 & \htwo &  1-0 Q(4) &  82.0  &  12.5&\\
24547.4 & \htwo  & 1-0 Q(5) & 166.2  &  12.5&\\
24755.4 & \htwo &  1-0 Q(6) &  90.0  &  12.5&\\
22952.9 & CO & 2-0          & 671.2  &   3.2&\\
23245.0 & CO & 3-1          & 720.2  &  12.5&\\
23534.6 & CO & 4-2          & 694.0  &  12.5&\\
23834.4 & CO&  5-3          & 471.5  &  12.5&\\
\end{longtable}

\onecolumn
\begin{longtable}{clccccccccc}
\caption[]{Forbidden lines observed along the jet}\\
\hline
                 &     &       &       &         & \multicolumn{2}{c}{Blue 1} &  \multicolumn{2}{c}{Blue 2} & \multicolumn{2}{c}{Blue 3} \\ 
$\lambda _{air}$ & Ion & Upper & Lower &$E_{up}$ &  Flux & $\Delta$(Flux) & Flux & $\Delta$(Flux) & Flux & $\Delta$(Flux) \\
(\AA) &  & level & level & (cm$^{-1}$) &  \multicolumn{6}{c}{(10$^{-17}$\,erg\,s$^{-1}$\,cm$^{-2}$)}  \\[+5pt]  
\endfirsthead
\caption{Continued.} \\
\hline
$\lambda _{air}$ & Ion & Upper & Lower &$E_{up}$ &  Flux & $\Delta$(Flux) & Flux & $\Delta$(Flux) & Flux & $\Delta$(Flux) \\
(\AA) &  & level & level & (cm$^{-1}$) &  \multicolumn{6}{c}{(10$^{-17}$\,erg\,s$^{-1}$\,cm$^{-2}$)}  \\[+5pt]  
\endhead
\hline\\[-5pt] 
\multicolumn{9}{c}{Oxygen} \\
\hline\\[-5pt] 
6300.3 & \oi   & $^1\!D_{2}$   & $^3\!P_{2}$   & 15867.8 &  16.0 & 0.09 & 9.3 & 0.09 & 8.1 & 0.09\\
6363.7 & \oi   & $^1\!D_{2}$   & $^3\!P_{1}$   & 15867.8 &  5.2  & 0.09 & 3.3 & 0.09 & 2.1 & 0.09\\
\hline\\[-5pt] 
\multicolumn{9}{c}{Sulphur} \\
\hline\\[-5pt] 
6716.4 & \sii & $^2\!D_{5/2}$ & $^4\!S_{3/2}$ & 14884.7   &   9.9 & 0.06 & 11.6 & 0.06 & 12.9 & 0.07\\
6730.8 & \sii & $^2\!D_{3/2}$ & $^4\!S_{3/2}$ & 14852.9   &  21.0 & 0.06 & 17.1 & 0.06 & 15.2 & 0.07\\
10286.7 &  \sii & $^2\!P_{3/2}$ & $^2\!D_{3/2}$ & 24571.5 &  9.9 & 0.3 & 3.3 & 0.4 & ... & ... \\
10320.5 &  \sii & $^2\!P_{3/2}$ & $^2\!D_{5/2}$ & 24571.5 &  12.1 & 0.3 & 4.8 & 0.4& ... & ... \\
10336.4 &  \sii & $^2\!P_{1/2}$ & $^2\!D_{3/2}$ & 24524.8 &  10.2 & 0.3 & 1.9 & 0.4& ... & ... \\
10370.5 &  \sii & $^2\!P_{1/2}$ & $^2\!D_{5/2}$ & 24524.8 &  4.8 & 0.3 & ... & ... & ... & ... \\
\hline\\[-5pt] 
\multicolumn{9}{c}{Nitrogen} \\
\hline\\[-5pt] 
6548.0 & \nii & $^1\!D_{2}$ & $^3\!P_{1}$ & 15316.1                &  0.9 & 0.06 & ... & ...& ... & ... \\
6583.4 & \nii & $^1\!D_{2}$ & $^3\!P_{2}$ & 15316.1                & 2.0 & 0.06 & 0.65 & 0.06& ... & ... \\
10397.7 & [\ion{N}{i}]  & $^2\!P_{3/2}$ & $^2\!D_{5/2}$ & 28839.3  & 10.4 & 0.3 & 3.4 & 0.4& ... & ... \\
10398.1     &      & $^2\!P_{1/2}$ & $^2\!D_{5/2}$ & 28838.9       &       &    \\
10407.1 & [\ion{N}{i}]  & $^2\!P_{3/2}$ & $^2\!D_{3/2}$ & 28839.3  & 8.6 &  0.3 & 3.8 & 0.04& ... & ... \\
10407.5     &      & $^2\!P_{1/2}$ & $^2\!D_{3/2}$ &  28838.9      &       &  \\
\hline\\[-5pt] 
\multicolumn{9}{c}{Iron} \\
\hline\\[-5pt] 
7155.1 & \feii & $a^2\!G_{9/2}$ & $a^4\!F_{9/2}$ & 15844.6   &  8.8 & 0.7 & 3.6 & 0.07 & 2.4 & 0.07\\
7172.0 & \feii & $a^2\!G_{7/2}$ & $a^4\!F_{7/2}$ & 16369.4   &   2.4  &   0.07 & 0.6 & 0.07 & ... & ... \\
7388.1 & \feii & $a^2\!G_{7/2}$ & $a^4\!F_{5/2}$ & 16369.4   &   2.3  &   0.07 & 0.6 & 0.07 & 0.6 & 0.08\\
7452.5 & \feii & $a^2\!G_{9/2}$ & $a^4\!F_{7/2}$ & 15844.65  &  3.5   &   0.07 & 1.2 & 0.07 & 0.9 & 0.08\\
7686.9 & \feii & $a^4\!P_{3/2}$ & $a^6\!D_{5/2}$ & 13673.2   &   1.5  &   0.08 & 0.6 & 0.06 & 0.5 & 0.08\\ 
8616.9 & \feii & $a^4\!P_{5/2}$ & $a^4\!F_{9/2}$ & 13474.4   &  20.8  &   0.1  & 9.0 &0.06 & 5.8 & 0.08\\
8891.9 & \feii & $a^4\!P_{3/2}$ & $a^4\!F_{7/2}$ & 13673.2   &   8.6  &   0.1  & 3.6 & 0.06 & 1.9 & 0.08\\
9033.4 & \feii & $a^4\!P_{1/2}$ & $a^4\!F_{5/2}$ & 13904.8   &   2.6  &   0.08 & 1.0 & 0.07 & 0.5 & 0.07\\
9051.9 & \feii & $a^4\!P_{5/2}$ & $a^4\!F_{7/2}$ & 13474.4   &   6.3  &   0.08 & 2.3 & 0.07 & 1.3 & 0.07\\
9226.6 & \feii & $a^4\!P_{3/2}$ & $a^4\!F_{5/2}$ & 13673.2   &   5.7  &   0.08 & 2.5 & 0.07 & 1.2 & 0.07\\
9267.5 & \feii & $a^4\!P_{1/2}$ & $a^4\!F_{3/2}$ & 13904.8   &   4.8  &   0.08 & 1.8 & 0.07 & 1.0 & 0.07\\
9399.0 & \feii & $a^4\!P_{5/2}$ & $a^4\!F_{5/2}$ & 13474.4   &   1.8  &   0.2  & 0.7 & 0.2  & 0.5 & 0.2\\
12485.4 & \feii & $a^4\!D_{5/2}$ & $a^6\!D_{7/2}$ & 8391.9   &   7.3  &   0.1 & 3.8 & 0.1 & 1.9 & 0.2\\
12521.3 & \feii & $a^4\!D_{1/2}$ & $a^6\!D_{3/2}$ & 8846.8   &   4.3  &   0.1 & 1.8 & 0.1 & ... & ... \\
12566.8 & \feii & $a^4\!D_{7/2}$ & $a^6\!D_{9/2}$ & 7955.3   & 313.0  &   0.1 & 171.1 & 0.1 & 127.2 & 0.6\\
12703.4 & \feii & $a^4\!D_{1/2}$ & $a^6\!D_{1/2}$ & 8846.8   &  23.7  &   0.5 & 8.7 & 0.5 & 6.9 & 0.6\\
12787.7 & \feii & $a^4\!D_{3/2}$ & $a^6\!D_{3/2}$ & 8680.45  &  41.4  &   0.5 & 14.5 & 0.5 & 7.7 & 0.4\\
12942.7 & \feii & $a^4\!D_{5/2}$ & $a^6\!D_{5/2}$ & 8391.94  &  52.6  &   0.2 & 20.5 & 0.2 & 10.1 & 0.4\\
12977.7 & \feii & $a^4\!D_{3/2}$ & $a^6\!D_{1/2}$ & 8680.45  &  14.7  &   0.2 & 5.6 & 0.2 &  ... & ... \\
13205.5 & \feii & $a^4\!D_{7/2}$ & $a^6\!D_{7/2}$ & 7955.30  & 107.7  &   0.2 & 57.1 & 0.4 & 42.1 & 0.4\\
13277.1 & \feii & $a^4\!D_{5/2}$ & $a^6\!D_{3/2}$ & 8391.94  &  28.2  &   0.2 & 10.8 & 0.4 & 5.7 & 0.4\\
15334.7 & \feii & $a^4\!D_{5/2}$ & $a^4\!F_{9/2}$ & 8391.94  &  80.5  &   0.1 &  9.2 & 0.1 & 14.0 & 0.1\\
15994.7 & \feii & $a^4\!D_{3/2}$ & $a^4\!F_{7/2}$ & 8680.45  &  65.4  &   0.2 & 21.8 & 0.1 & 8.0 & 0.1\\
16435.4 & \feii & $a^4\!D_{7/2}$ & $a^4\!F_{9/2}$ & 7955.3   & 527.1  &   0.3 & 258.5 & 0.2 & 175.9 & 0.6\\
16637.6 & \feii & $a^4\!D_{1/2}$ & $a^4\!F_{5/2}$ & 8846.8   &  38.6  &  0.3 & 13.8 & 0.2 & 6.4 & 0.6\\
16768.7 & \feii & $a^4\!D_{5/2}$ & $a^4\!F_{7/2}$ & 8391.9   & 76.1  &   0.3 & 26.2 & 0.2 & 10.6 & 0.6\\
17111.3 & \feii & $a^4\!D_{3/2}$ & $a^4\!F_{5/2}$ & 8680.4   & 18.3  &   0.1 & 7.0 & 0.1 & 3.4 & 0.2\\
17484.2 & \feii & $a^4\!P_{3/2}$ & $a^4\!D_{7/2}$ & 7955.30  & 4.5  &   0.3 & 1.8 & 0.1 & ... & ...\\
17971.1 & \feii & $a^4\!D_{3/2}$ & $a^4\!F_{3/2}$ & 8680.4   & 54.7  &  0.1 & 14.5 & 0.2 & 6.0 & 0.2\\
18000.2 & \feii & $a^4\!D_{5/2}$ & $a^4\!F_{5/2}$ & 8391.9   & 66.3  &  0.1 & 26.7 & 0.6 & 12.1 & 0.2\\
18093.9 & \feii & $a^4\!D_{7/2}$ & $a^4\!F_{5/2}$ & 7955.3   & 134.1  & 0.3 & 60.4 & 0.7 & 39.2 & 0.4\\
18954.0 & \feii & $a^4\!D_{5/2}$ & $a^4\!F_{3/2}$ & 8391.9   &  13.2  &  0.2 & 4.2 & 0.7 &... & ... \\
20151.2 & \feii & $a^2\!H_{9/2}$ & $a^2\!G_{9/2}$ & 20805.8  &  15.8  &  0.8 & ... & ...&... & ... \\
20460.1 & \feii & $a^2\!P_{3/2}$ & $a^4\!P_{5/2}$ & 18360.6  &   7.9  &  0.3 & ... & ...&... & ... \\
21327.7 & \feii & $a^2\!P_{3/2}$ & $a^4\!P_{3/2}$ & 18360.6  &   4.6  &  0.4  &... & ...&... & ... \\
22237.6 & \feii & $a^2\!H_{11/2}$ & $a^2\!G_{9/2}$ & 20340.3 &  21.7  &  0.4 & 7.3 & 0.7&... & ... \\
\hline\\[-5pt] 
\multicolumn{9}{c}{Calcium} \\
\hline\\[-5pt] 
7291.4 & \caii & $^2\!D_{5/2}$ & $^2\!S_{1/2}$ & 13710.8            &  11.9  &   0.07 & 6.7 & 0.07 & 6.5 & 0.07\\
7323.8 & \caii & $^2\!D_{3/2}$ & $^2\!S_{1/2}$ & 13650.1            &   7.7 &    0.07 & 4.5 & 0.07 & 3.9 & 0.08\\
8542.0$^a$ & \ion{Ca}{ii} & $^2\!P_{3/2}$ & $^2\!D_{5/2}$ & 25414.4     & 1.3 & 0.07 & 3.7 & 0.06 & 2.5 & 0.09\\
\hline\\[-5pt] 
\multicolumn{9}{c}{Other forbidden lines} \\
\hline\\[-5pt] 
7377.8 & \niii & $^2\!F_{7/2}$ & $^2\!D_{5/2}$ & 13550.3     &  5.7  &   0.07 & 3.1 & 0.07 & 6.6 & 0.07\\
7411.6 & \niii & $^2\!F_{5/2}$ & $^2\!D_{3/2}$ & 14995.5     &   0.9 &   0.07 & ... & ...& ... & ...\\
19387.7 & \niii & $^2\!F_{7/2}$ & $^4\!F_{9/2}$ & 13550.3    &  40.4  &  0.2  & 10.3 & 0.07 & ... & ...\\
8125.3 & \crii & $a^6\!D_{7/2}$ & $a^6\!S_{5/2}$ & 12303.8   &   1.0  &  0.08 & 0.9 & 0.06 & 4.7 & 0.08\\
8308.4 & \crii & $a^6\!D_{3/2}$ & $a^6\!S_{5/2}$ & 12032.5   &   0.9  &  0.06 & ... & ...& ... & ...\\
8357.6 & \crii & $a^6\!D_{1/2}$ & $a^6\!S_{5/2}$ & 11961.8   &   0.5  &  0.06 & ... & ...& ... & ...\\
8729.9 & \crii & $a^4\!F_{3/2}$ & $a^4\!D_{3/2}$ & 31082.9   &   0.8  &  0.1  & 1.1 & 0.06 & 0.98 & 0.08\\
9824.1 & [\ion{C}{i}] &  $^1\!D_{2}$ & $^3\!P_{1}$ & 10192.6 &   8.1  &  0.2 & 11.9 & 0.2 & 13.9 & 0.2\\
9850.2 & [\ion{C}{i}] &  $^1\!D_{2}$ & $^3\!P_{2}$ & 10192.6 &  20.7  &  0.2 & 31.1 & 0.2 & 47.9 & 0.2\\
\hline\\[-5pt] 
\end{longtable}

\end{appendix}
\end{document}